\documentclass{emulateapj}

\newcommand{\kms}{\mbox{km s$^{-1}$}}
\newcommand{\msun}{\mbox{M$_{\odot}$}}
\newcommand{\msunyr}{\mbox{M$_{\odot}$yr$^{-1}$}}

\shorttitle{Kinematic Structure of Merger Remnants}
\shortauthors{Cox, et al.}

\begin{document}

\title{The Kinematic Structure of Merger Remnants}

\author{T. J. Cox\altaffilmark{1},
Suvendra N. Dutta\altaffilmark{1},
Tiziana Di Matteo\altaffilmark{2},
Lars Hernquist\altaffilmark{1},
Philip F. Hopkins\altaffilmark{1},
Brant Robertson\altaffilmark{1},
and Volker Springel\altaffilmark{3}}

\altaffiltext{1}{Harvard-Smithsonian Center for Astrophysics,
60 Garden Street, Cambridge, MA 02138, USA}
\altaffiltext{2}{Carnegie-Mellon University, Dept. of Physics,
5000 Forbes Ave., Pittsburgh, PA 15213, USA}
\altaffiltext{3}{Max-Planck-Institut f\"{u}r Astrophysik,
Karl-Schwarzchild-Stra\ss e 1, 85740 Garching bei M\"{u}nchen, Germany}

\begin{abstract}

We use numerical simulations to study the kinematic structure of
remnants formed from mergers of equal-mass disk galaxies.  
In particular, we show
that remnants of dissipational mergers, which include the radiative
cooling of gas, star formation, feedback from supernovae, and the
growth of supermassive black holes, are smaller, rounder, have, on
average, a larger central velocity dispersion, and show significant
rotation compared to remnants of dissipationless mergers.  The 
increased rotation speed of dissipational remnants owes its origin to
star formation that occurs in the central regions during the galaxy
merger.  We have further quantified the anisotropy, three-dimensional
shape, minor axis rotation, and isophotal shape of each merger remnant,
finding that dissipational remnants are more isotropic, closer to
oblate, have the majority of their rotation along their major axis, 
and are more disky than dissipationless remnants.  Individual remnants
display a wide variety of kinematic properties.  A large fraction of
the dissipational remnants are oblate isotropic rotators.  Many
dissipational, and all of the dissipationless, are slowly rotating
and anisotropic.  The remnants of gas-rich major mergers can
well-reproduce the observed distribution of projected ellipticities,
rotation parameter ($V/\sigma$)$^*$, kinematic misalignments, $\Psi$,
and isophotal shapes.  The dissipationless remnants are a poor match
to this data.  We also investigate the properties of merger remnants
as a function of initial disk gas fraction, orbital angular momentum,
and the mass of the progenitor galaxies.  Our results support the
merger hypothesis for the origin of low-luminosity elliptical galaxies 
provided that the progenitor disks are sufficiently gas-rich, however
our remnants are a poor match to the bright ellipticals that are
slowly rotating and uniformly boxy.  

\end{abstract}

\keywords{galaxies: ellipticals --- evolution --- formation --- interactions 
          --- methods:N-body simulations}

\section{Introduction}
\label{sec:intro}

The observed absorption-line spectra and red colors of elliptical
galaxies suggest that their stars were formed at high
redshift ($z\geq1$) and that very little star formation has occurred
in them since then.  According to the ``merger hypothesis''
\citep{TT72,T77}, these red elliptical galaxies are produced by the
collision and merger of spiral galaxies, and hence the progenitors of
present day ellipticals may be high-redshift spirals.  While
relatively little is known about disk galaxies at high redshift, it is
likely that these disks were more concentrated and gas-rich than their
low-redshift counterparts.  Indeed, preliminary observational evidence
\citep{Erb06m} indicates that galaxies at
redshift $z\approx 2$ do have large gas fractions $f_{\rm gas}\sim0.5$,
with some approaching $f_{\rm gas}\sim0.8 - 0.9$.  Thus, any attempt to
understand the formation, properties and scaling relations of the
present day population of elliptical galaxies, within the context of
the ``merger hypothesis'', must consider gas-rich mergers.

Requiring that the disk galaxy progenitors contain a significant
fraction of gas is nothing new to the ``merger hypothesis''.  One of
the main objections to this mechanism of producing ellipticals, argued
by, for instance, \citet{Ost80}, is that the centers of ellipticals
are more concentrated than local spirals.  Cast in terms of
phase-space density, this objection states that the high central
phase-space density of ellipticals cannot be produced by the merger of
low phase-space spirals because, according to Liouville's Theorem,
phase-space density is conserved during a collisionless process
\citep{Car86}.  However, this argument breaks down when the merger
constituents contain gas, which can radiate energy, and hence increase
the phase space density \citep{Lak89}.  An estimate of how much gas is
required to match the central densities of ellipticals was provided by
\citet{HRemIII}, who used N-body simulations and analytic arguments to
suggest that mergers of spiral galaxies containing $\geq30$\% gas
would be sufficient to account for the high phase space densities of
ellipticals.

Within the context of the hierarchical theory of structure formation,
gas-rich major mergers may play a much larger role than just resolving
the central phase-space densities of elliptical galaxies.  Previously,
we have described a ``cosmic cycle'' of galaxy formation and evolution
in which gas-rich mergers drive the evolution of quasars
\citep{Hop06big}, induce the growth of supermassive black holes
\citep{dMSH05}, and produce red elliptical galaxies
\citep{SdMH05red,Hop06red} that obey many of the observed scaling
relations \citep{Rob06fp,Rob06b}.  A schematic view of this picture is
presented in Fig. 1 of \citet{Hop06big}.  However, the success of this
scenario also must be gauged by its ability to produce remnants that
have kinematic and morphological properties characteristic of observed
ellipticals.  It is this question that we address in the current
paper.

Observations indicate that galaxy spheroids can be classified into two
groups \citep[][and references therein]{Dav83,Ben89,Ben88,
Fab97,KB96}.  Large, luminous spheroids have hot gaseous halos,
box-shaped isophotes, surface-brightness profiles with flat ``cores,''
show very little rotation and are almost uniformly classified as
ellipticals.  Less luminous spheroids tend to have little, if any, hot
gas, disk-shaped isophotes, power-law surface-brightness profiles, and
exhibit rotation that is along the photometric major axis.  The
latter group encompasses many low-luminosity ellipticals, bulges and
S0s.

The last of these properties, the alignment and rotation of spheroidal
galaxies, may be the result of a more fundamental dichotomy among
elliptical galaxies; that is, the isotropy of their velocity
distributions.  Because spheroids can be flattened for (at least) two
reasons, rotation and velocity anisotropy, the lack of rotation in
large ellipticals suggests that these systems have significant
velocity anisotropy, while many low-luminosity ellipticals and bulges
are consistent with being isotropic systems flattened by their
observed rotation.

One viewpoint is that these two classes of elliptical galaxies exhibit
different properties because they are formed via different mechanisms.
Along this line of reasoning, \citet{NB03} argue that large
ellipticals are formed by the dissipationless merger of two
comparable-mass disk galaxies, i.e., major mergers, while
low-luminosity ellipticals are produced by the dissipationless merger
of unequal mass disks, i.e., minor mergers.  To demonstrate this
possibility, \citet{NB03} used numerical simulations of
dissipationless disk galaxy mergers to show that remnants of major
mergers rotate very little and are, in general, boxy, similar to
luminous ellipticals.  On the other hand, the simulated minor mergers
rotate significantly and have disky isophotes, similar to
low-luminosity ellipticals.  However, these simulations did not
include gas physics, and hence would not induce starbursts, quasar
activity, nor satisfy the scaling relations of elliptical galaxies
\citep{Rob06fp}.  Moreover, if most ellipticals are relatively old,
i.e. have mainly old stellar populations, and were formed long ago by
mergers, it is likely that the progenitor galaxies would have been
gas-rich and had a higher gas fraction than local large, star-forming
galaxies.

In this paper we use a large suite of numerical simulations to explore
the kinematic properties of remnants produced by the merger of
comparable-mass disk galaxies.  We specifically address the
differences between remnants formed via dissipationless mergers versus
those produced in gas-rich mergers that include the cooling of gas,
star formation and feedback.  We find that gas-rich mergers can
successfully reproduce many of the kinematic properties of observed
elliptical galaxies, while dissipationless remnants provide a poor
match the data.

The organization of the rest of this paper is as follows.  In
\S~\ref{sec:meth} we describe the numerical simulations,
including the disk galaxy models and the galaxy collisions
(\ref{ssec:sims}), followed by the techniques employed to analyze
individual merger remnants (\ref{ssec:anal}).  In
\S~\ref{sec:results} we present the results of our analysis for
the entire series of merger remnants.  To begin, we report the
aggregate of measured properties for our entire series of simulated
merger remnants (\ref{ssec:basics}).  Following this, we show the
remnant rotational support (\ref{ssec:majrot}), including where this
rotation originates (\ref{ssec:origin}), and what individual remnants
are like (\ref{ssec:ind}).  To better characterize the remnants we
also analyze their anisotropy (\ref{ssec:anisot}), shape
(\ref{ssec:shapes}), minor-axis kinematics (\ref{ssec:minrot}), and
how these quantities depend on our various input assumptions
(\ref{ssec:dep}).  Finally, in \S~\ref{sec:disc} we discuss the
implications of our results for the formation of elliptical galaxies
and we conclude in \S~\ref{sec:conclusions}.

\section{Methods}
\label{sec:meth}

%
\subsection{Merger Simulations}
\label{ssec:sims}

The simulations presented here are part of a large, ongoing effort to
study galaxy mergers and how this process impacts the formation and
evolution of galaxies.  Hence, the methods used here are identical to,
and also described in, several related works
\citep{Cox06x,dMSH05,Hop05b,Hop06big,Hop05c,Hop05d,Hop06slope,
Hop06red,Hop05a,Lidz06,Rob06fp,Rob05a,Rob06b,SdMH05red,SdMH05}.
Readers that desire a more detailed description of the simulation
code and the construction of the progenitor galaxy models are referred
to \citet{SdMH05}, where the majority of the methods were first
introduced.  Below, we provide a brief overview of our methodology and
focus on the assumptions most relevant for the results reported here.

The simulations described here were performed using the N-body/SPH
(smoothed particle hydrodynamics) code {\small GADGET2}
\citep{SpGad2}, which is based on a fully conservative formulation of
SPH \citep{SHEnt}.  Along with the standard features of this
publically available code, the version we employ includes radiative
cooling of gas, star formation that is designed to match the observed
Schmidt-law \citep{Sch59,Kenn98}, the multiphase feedback model of
\citet{SH03} softened ($q_{\rm EOS}$=0.25) so that the
mass-weighted interstellar medium temperature is $\sim10^{4.5}$~K,
and a centrally located sink particle that can accrete
gas and release isotropic thermal energy that represents a massive
black hole \citep{SdMH05}.

The fiducial galaxy model consists of a dark matter halo and an
embedded rotationally-supported exponential disk.  For simplicity we
do not include a spheroidal bulge in our fiducial galaxy model.
The dark matter
halo is initialized with a \citet{H90} profile that has an effective
concentration of 9, a spin parameter $\lambda=0.033$, and a circular 
velocity $V_{200}=160$~\kms.  The exponential disk size is fixed by
requiring that the disk mass fraction (4.1\% of the total mass) is
equal to the disk spin fraction.  This results in a radial disk scale
length of 3.9~kpc.  In addition, a specified fraction $f$ of the disk
mass is in a gaseous component.  The galaxy models are realized by
120,000 dark matter particles, and 80,000 disk particles.  A fraction
$f\times80,000$ of the disk particles represent the gaseous disk, and
the remainder represent the collisionless stellar disk.

In the present study, we restrict our analysis to simulations of major
mergers between identical disk galaxies.  We also consider only two
types of progenitor disks.  One, where the exponential disk is purely
stellar ($f=0$), and a second where the disk is 40\% gas ($f=0.4$).
In all simulations, we adopt a gravitational softening length of 
140~pc.  We restrict our analysis to quantities measured on the scales
of the effective radius, i.e., several kpc, where we are confident that
we have sufficient resolution.  Unfortunately, with this resolution we
cannot reliably determine the inner ($\leq 140$~pc) surface density
profiles \citep[][and references therein]{Lau05} nor follow the merger
of the binary black holes \citep[see, e.g.,][]{Mil01}.

Once built, pairs of identical galaxies are placed on parabolic orbits
with the spin axis of each disk specified by the angles $\theta$, and
$\phi$ in standard spherical coordinates.  Table~\ref{tab:orbs} lists
the adopted orientation of each progenitor for our fifteen simulated
mergers.  The orientation list contains seven idealized mergers
(labeled $b-h$) that represent orientations often seen in the
literature.  For example, orientation $h$ has the galactic spin and
orbital angular momentum aligned in what is typically called a
prograde-prograde merger.  Also included are slight variations of this
case, e.g., prograde-retrograde, retrograde-retrograde, and polar.  
The remainder of the mergers (labeled $i-o$) follow \citet{B92}
in that they are selected to be unbiased initial disk orientations
according to the coordinates of two oppositely directed tetrahedra.
These orbits are identical to those explored by e.g. \citet{NB03} and
therefore allow for a direct comparison to their results.

\begin{deluxetable}{lccccl}
\tabletypesize{\scriptsize}
\tablecaption{Disk Orientations\label{tab:orbs}}
\tablewidth{0pt}
\tablehead{
\colhead{Run} &
\colhead{$\theta_1$} &
\colhead{$\phi_1$} &
\colhead{$\theta_2$} &
\colhead{$\phi_2$} &
\colhead{Comments} \\
\colhead{(1)} & \colhead{(2)} & \colhead{(3)} &
\colhead{(4)} & \colhead{(5)} & \colhead{(6)}
}
\startdata
h &   0  &   0 &   0 &   0 & both prograde \\
b & 180  &   0 &   0 &   0 & prograde-retrograde \\
c & 180  &   0 & 180 &   0 & both retrograde \\
d &  90  &   0 &   0 &   0 & polar 1 \\
e &  30  &  60 & -30 &  45 & tilted 1 \\
f &  60  &  60 & 150 &   0 & polar 2 \\
g & 150  &   0 & -30 &  45 & tilted 2 \\
i &   0  &   0 &  71 &  30 & Barnes orientations \\
j & -109 &  90 &  71 &  90 &     $\Downarrow$ \\
k & -109 & -30 &  71 & -30 &      \\
l & -109 &  30 & 180 &   0 &   \\
m &    0 &   0 &  71 &  90 &   \\
n & -109 & -30 &  71 &  30 &   \\
o & -109 &  30 &  71 & -30 &   \\
p & -109 &  90 & 180 &   0 &   \\
\enddata
\tablecomments{List of disk galaxy orientations 
for major merger simulations.  Col. (1) is our
unique orientation identification. Cols. (2) and
(3) are the initial orientation of disk 1 and
cols. (4) and (5) are the orientation of disk 2.
A brief description of several unique orientations
is listed in column (6).
}
\end{deluxetable}

The primary results of our work are based upon two series of simulated
major mergers.  The first series consists of the fifteen orientations
listed in Table~\ref{tab:orbs} each simulated with progenitor disks
composed of purely collisionless stellar mass ($f=0$).  The second
series is identical to this first series but the progenitor disks all
contain 40\% gas ($f=0.4$).  In what follows we will consistently
refer to the former series as ``dissipationless'' mergers, while the
latter are ``dissipational'', or ``40\% gas'' mergers.

In \S~\ref{ssec:dep} we consider additional simulations to
address how robust our results are to the chosen initial conditions.

%
\subsection{Remnant Analysis}
\label{ssec:anal}

\begin{figure*}
\plotone{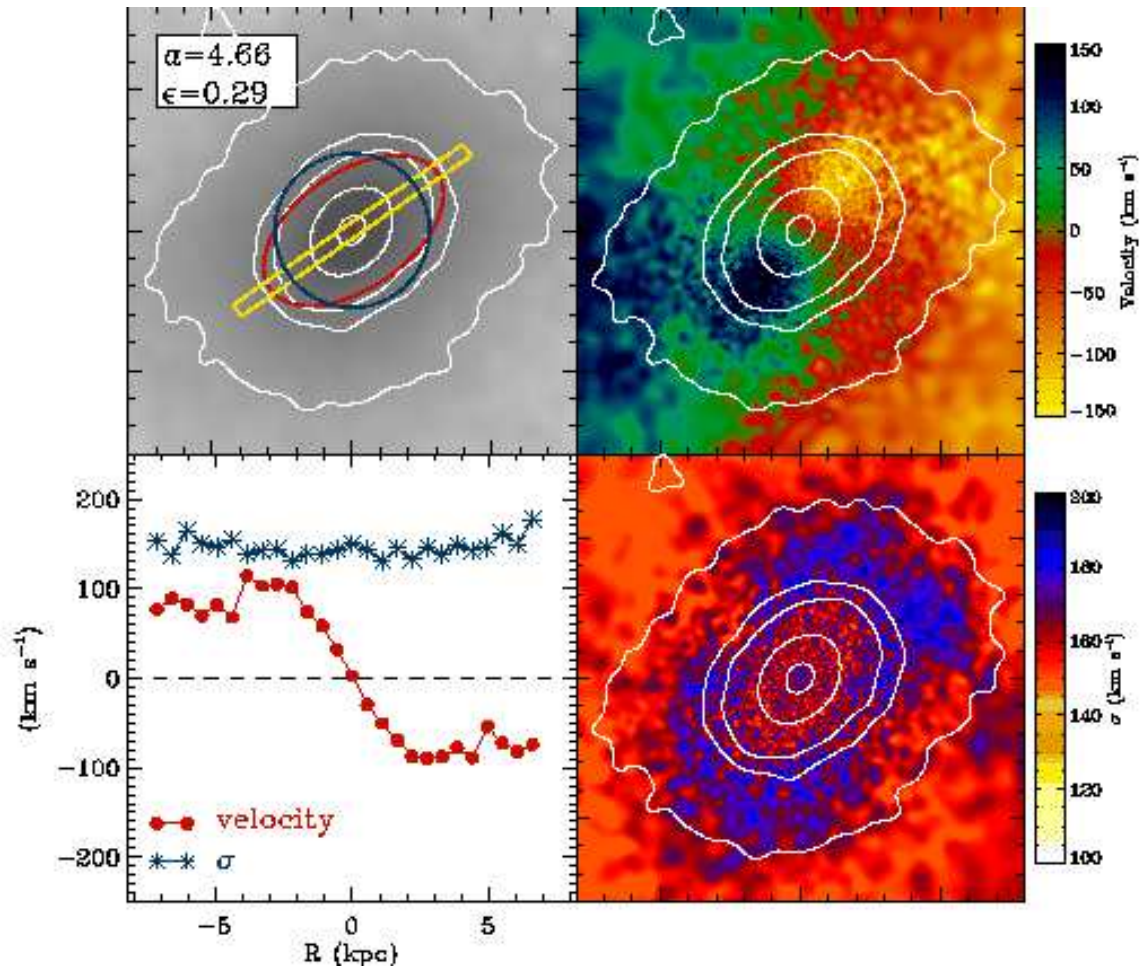}
\caption{Projected mass density ({\it upper left}) and velocity
({\it upper right}) and dispersion ({\it lower right}) fields
 for one of our merger remnants (dissipational merger, orientation $f$,
viewed perpendicular to the orbital plane)  with 
overlaid isodensity contours.  The half-mass radius, computed within
a circular aperture, is shown by
a ($blue$) circle.  The half-mass isodensity contour has been
fit with a ($red$) ellipse.  The semi-major axis length $a$ and
ellipticity $\epsilon$ are shown in the upper-left of the figure.
A slit has been placed along the major axis and the velocity along
the slit is show in the lower left panel.  This remnant is a 
fast rotator.
\label{fig:remimg}}
\end{figure*}

The techniques we employ to analyze the merger remnants are designed
to mirror those typically adopted by observers.  For each remnant we
project the stars onto a plane as if observed from a particular
direction.  An example of the projected stellar mass is shown in the
upper-left plot of Figure~\ref{fig:remimg}.  The stellar mass is
composed of two components.  One, the dissipationless disk particles
that are present in our initial conditions, and two, stars that are
formed during the simulation from dense gas, when included.

Once the stellar mass is projected onto a plane, we determine the
iso-density contour that encloses half of the stellar mass and fit an ellipse
to it.  We label the semi-major axis of this ellipse $a$, and the
semi-minor axis $b$.  The ellipticity is then defined as $\epsilon = 1
- b/a$.  We note that $a$ is slightly larger than the half-mass radius
$R_e$, which is typically calculated using a circular aperture.  Both
the circular-aperture half-mass radius, and the fitted half-mass ellipse
are overlaid on the projected surface density in
Figure~\ref{fig:remimg}.  In addition, the iso-density contours are
shown.  The remnant shown in Figure~\ref{fig:remimg} is a
dissipational merger from the $f$ orbit.  It is a fast rotator viewed
from an angle perpendicular to the orbital plane.

We note that observations often determine $R_e$ in a manner which differs
from our approach.  In that case, the surface brightness profile is
fit to an analytic function, e.g., a $R^{1/4}$ \citep{deV48} or Sersic
\citep{Ser68} profile, from which the effective radius is extracted.
Unfortunately, owing to the limited resolution of telescopes and finite
backgrounds, this technique is subject to uncertainties that depend on
the radial range over which the fit is performed and the assumed profile
\citep{Tru01}.  In fact, \citet{BK05} show that the $R_e$ of \citet{H90}
profile can be underestimated by nearly 30\% when the maximum radius used
for the fitting produces is varied between $\sim2$~$R_e$ to $>10$~$R_e$.
Because of these complications, and because we know the exact location
of all stellar material we can extract the half-mass radius directly from
the particle information.  This eliminates any ambiguity between the 
procedure used to fit the profile and the resultant $R_e$.  In what
follows we will typically use $a$, rather than $R_e$, as a measure of
the ``size'' of our merger remnants.

To quantify the kinematics of each remnant, we place a slit along the
major axis and measure the velocity and dispersion along the slit.
Again, this is demonstrated in Figure~\ref{fig:remimg} where we show
the projected two dimensional rotation and dispersion velocity fields,
the slit, and the resultant velocity profiles.  The slit has length
and width of $\sim3 a$, and $0.25 a$, respectively.  The slit is
divided into 26 bins, lengthwise, and the line of sight velocity field
is measured for each bin.  For simplicity, the velocity profile in
each bin is assumed to be Gaussian, and thus we extract the mean
rotation velocity and velocity dispersion from each bin.

The rotation velocity $V_{\rm maj}$ along the major axis is determined
as the average absolute value of the maximum and minimum velocity
along the slit.  The velocity dispersion $\sigma$ of each remnant is
the average of all dispersions $R<0.5a$ along the slit.  In practice,
the dispersion profile is fairly flat out to $\sim R_e$ and thus the
choice of 0.5 does not affect our results.  The large aperture used to
measure $\sigma$ is also selected becuase it is much greater than the
numerical resolution.

We note that the two dimensional velocity fields contain a wealth of
information that is largely lost by the use of a simple slit.
However, the majority of velocity data taken to date has been
obtained using
slits and we therefore follow this procedure in our work here.  We do
note, though, that a fruitful avenue for future work is the comparison
of simulated two dimensional velocity fields such as those presented
in Figure~\ref{fig:remimg} to observational samples such as SAURON
\citep{Cap06}, which use integral field spectroscopy and thus image
the two-dimensional velocity field.

\begin{figure}
\plotone{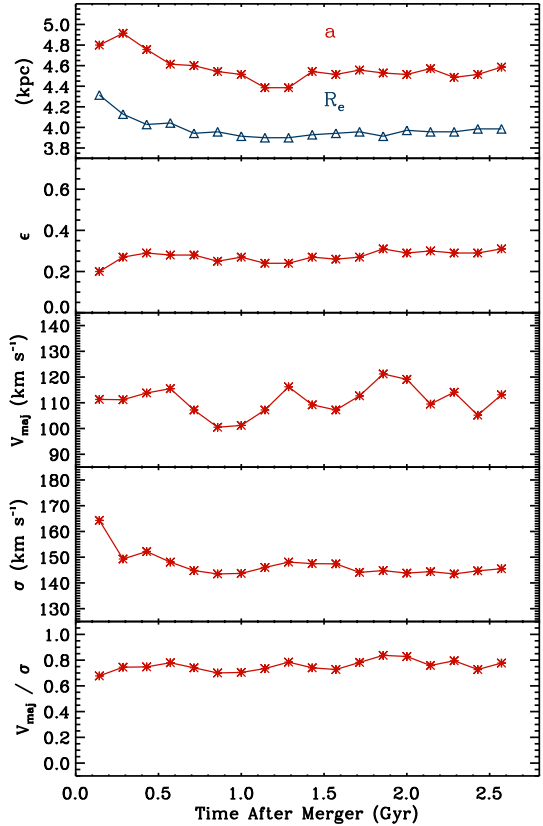}
\caption{Time evolution of the semi-major axis $a$,
half-mass radius $R_e$, ellipticity $\epsilon$, 
major axis rotation speed $V_{\rm maj}$, central
velocity dispersion $\sigma$, and the ratio
$V_{\rm maj}/\sigma$ for the same merger remnant
displayed in Figure~\ref{fig:remimg}.
\label{fig:f_time}}
\end{figure}

Because our models do not account for cosmological evolution, the
simulation time is not directly connected to redshift.  The initial
conditions are designed to approximate our own Milky Way galaxy
except, when included, a higher disk gas fraction.  Thus, the mergers
presented here are probably more representative of moderate redshifts
where disks were likely gas-rich.  In fact, if the Milky Way has been
forming stars at a steady rate of 1\msunyr, without any other mass flux,
it would have had a gas fraction above 40\% about 10 Gyr ago ($z\sim1.5$).
In any case, we will mainly compare our remnants
to local elliptical galaxies, most of which are old stellar systems and
dynamically relaxed.  Hence, we would like to ensure that the
merger remnants are relaxed systems as well.  In order to investigate
the dynamical stability of our remnants, Figure~\ref{fig:f_time} shows
the size, both $a$ and $R_e$, ellipticity $\epsilon$, major axis
rotation speed $V_{\rm maj}$, central velocity dispersion $\sigma$,
and the ratio $V_{\rm maj}/\sigma$ as a function of time after the
merger is complete for the same remnant that was shown in
Figure~\ref{fig:remimg}.
For the most part, the remnant relaxes quickly ($<300$~Myr) after the
merger.  Prior to this, the central dispersion $\sigma$, the rotation
speed, and size appear to be the most strongly time-dependent
quantities.  In
particular, the dispersion is about $\sim$15\% larger immediately
after the merger compared to the relaxed remnant.  Of all the
quantities, the rotation speed fluctuates the most, and it varies
within 10\% of its average value, throughout the
simulation.  These 10\% fluctuations are carried over to the ratio
$V_{\rm maj}/\sigma$, and thus we consider this a reasonable
quantification of the error associated with {\it when} we observe each
remnant, which we take to be $\sim2.5$~Gyr after the merger is
complete.

As a final comment, we note that all results presented in this work
use the stellar mass as opposed to the stellar light.  This choice is
motivated in part by the uncertain star formation history of all particles
designated as stars at the beginning of the simulation.  As we show in
\S~\ref{ssec:origin}, the stars formed during the simulation tend
to rotate more than those present at the beginning of the simulation,
and thus our choice to weight the velocity by mass as opposed to
luminosity may underestimate the velocities as compared to recent
merger remnants while being a fair representation of most old ellipticals.

\section{Results}
\label{sec:results}

%
\subsection{Effective Radius, Ellipticity,
Dispersion, and Rotation}
\label{ssec:basics}

\begin{figure*}
\plottwo{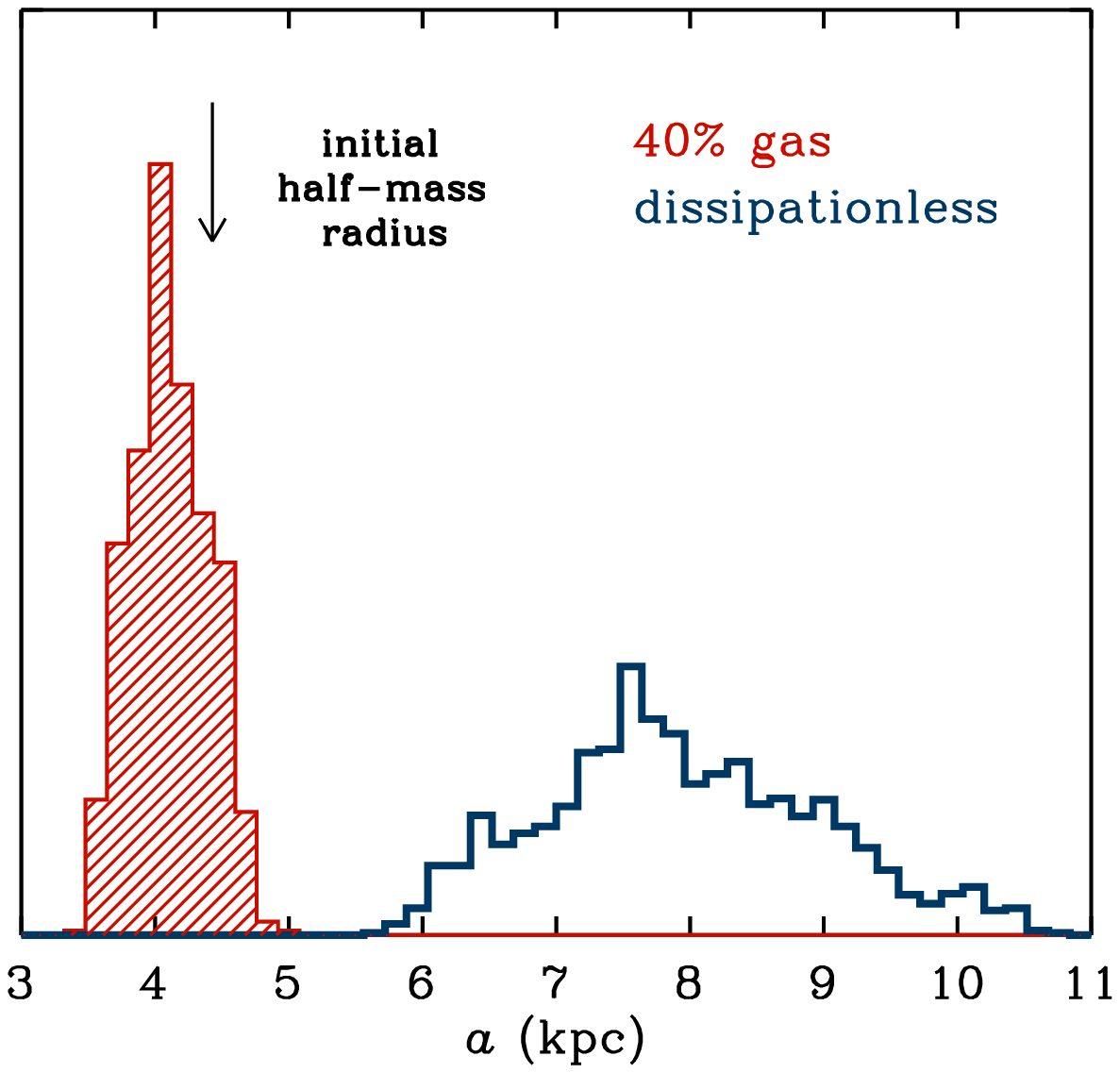}{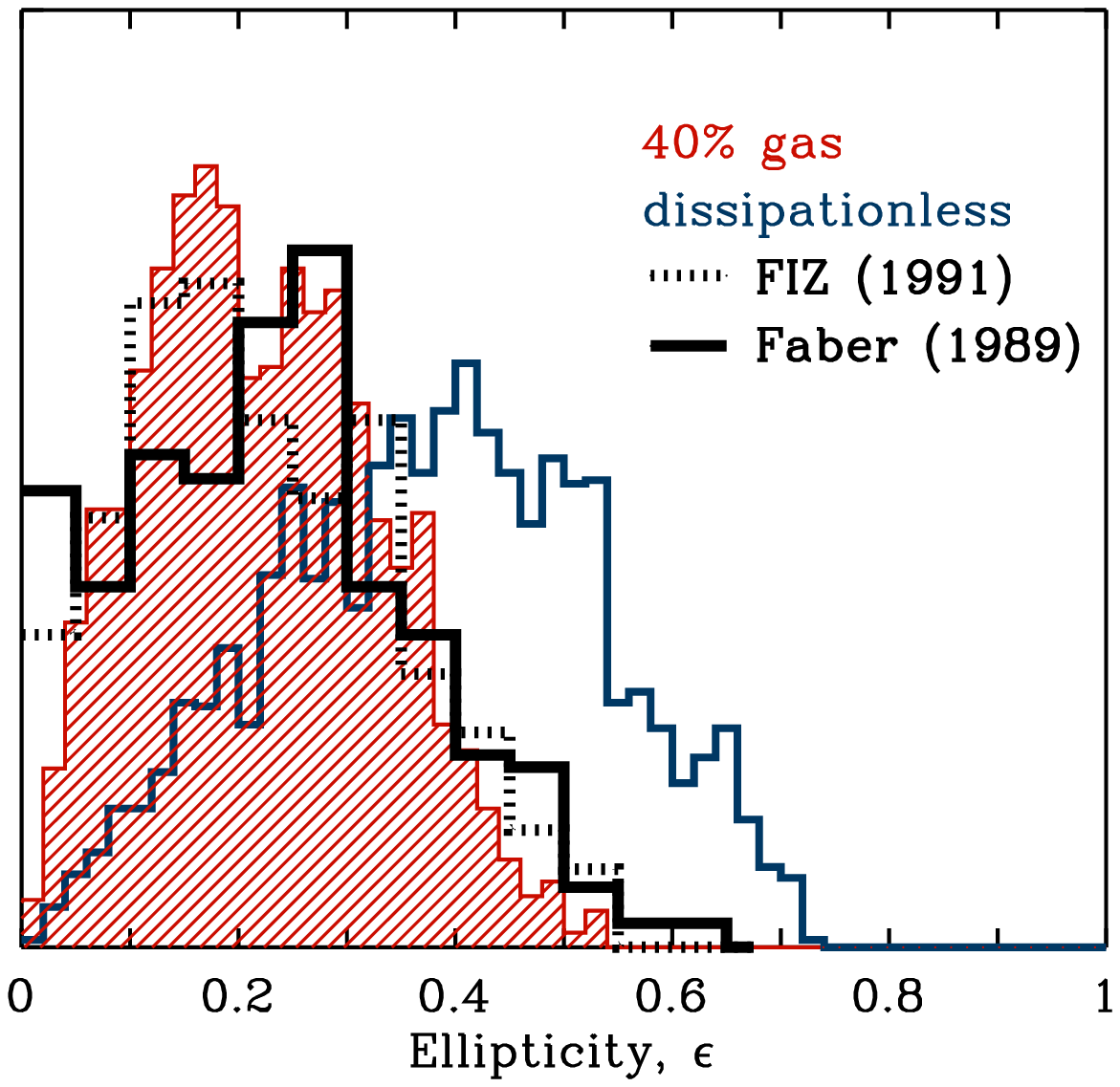}\\
\plottwo{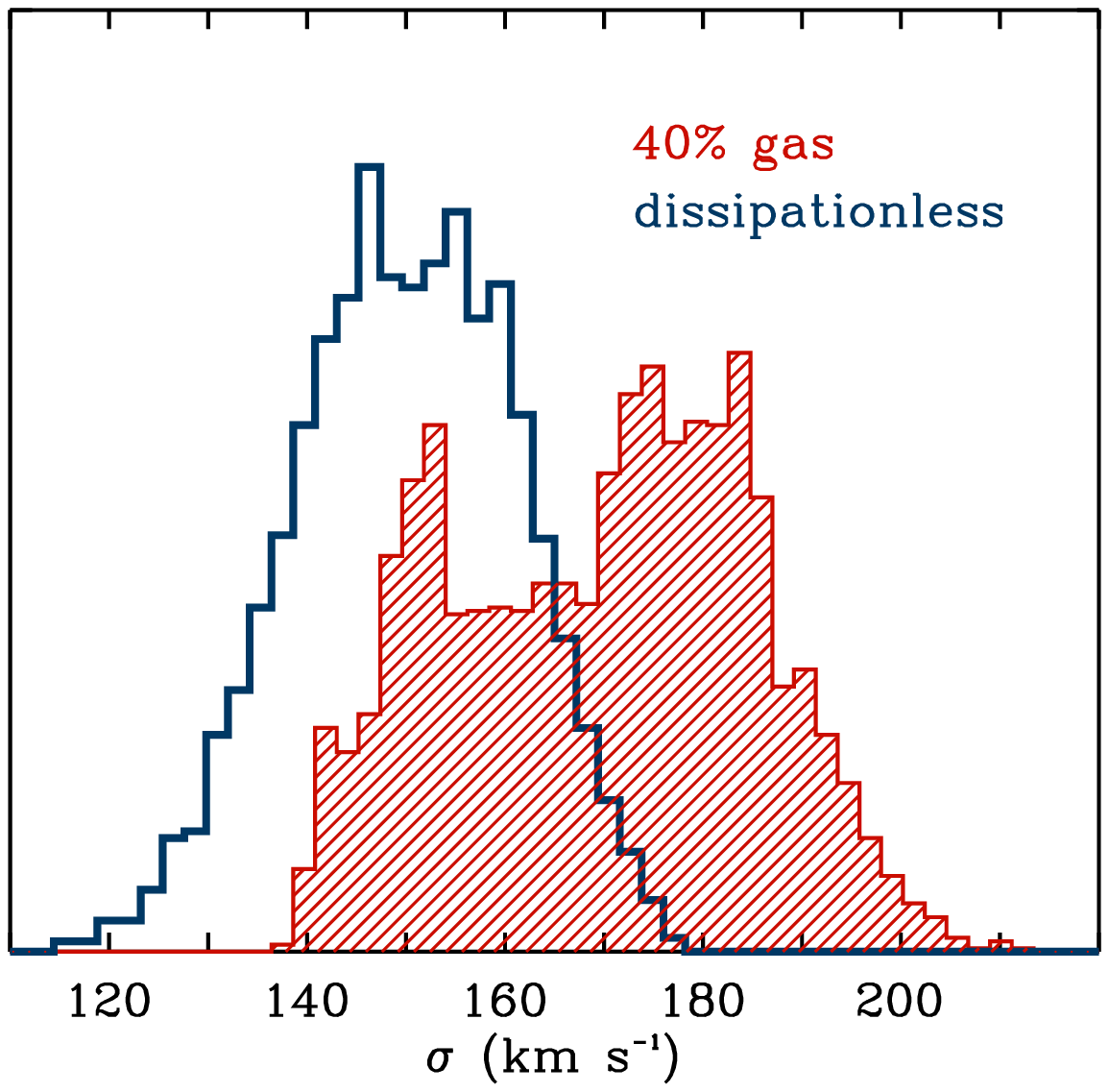}{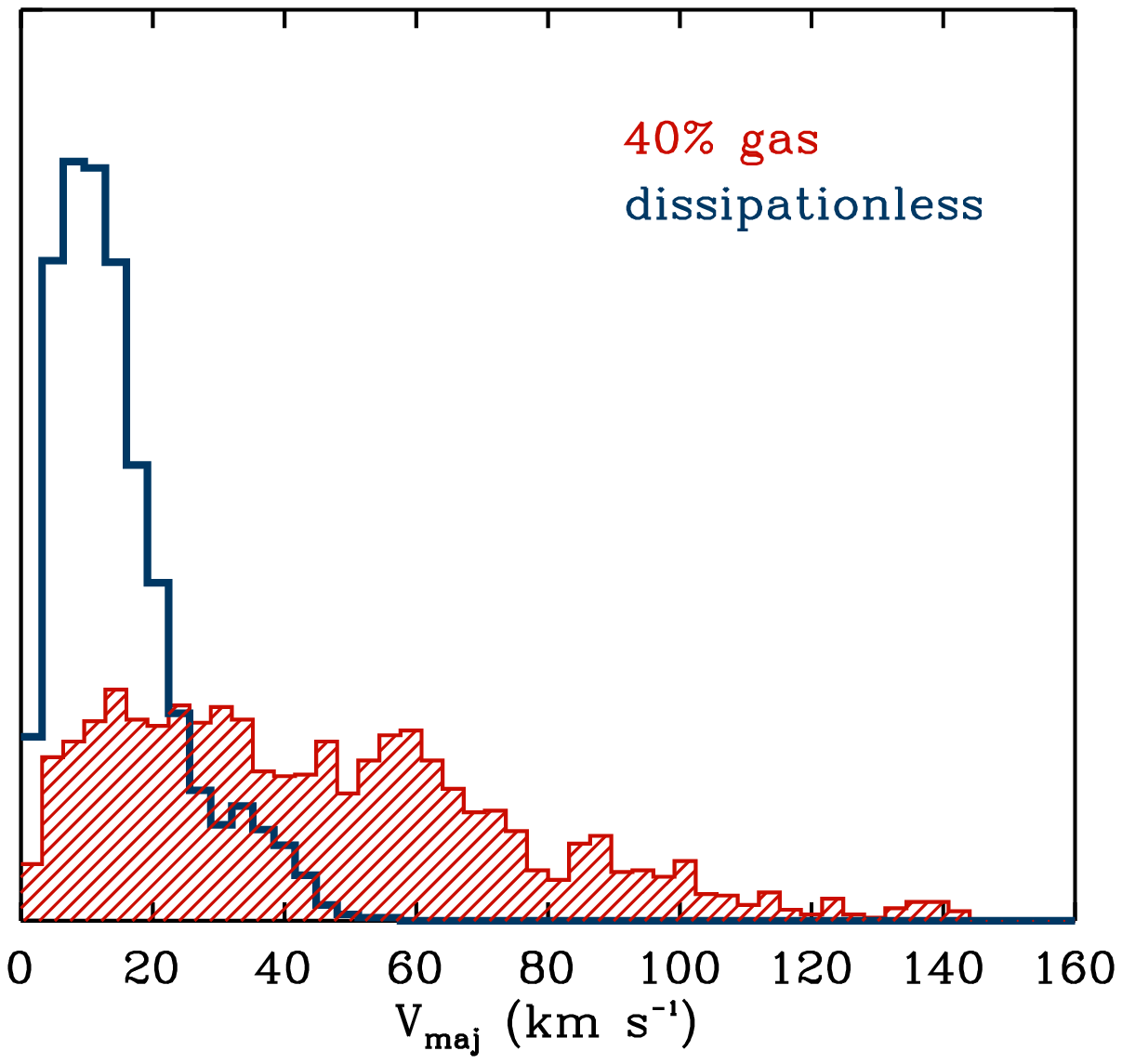}
\caption{Histograms of remnant properties.  Shown in ($red$)
cross-hatch are the dissipational remnants, while the open
($blue$) histogram is for the dissipationless remnants.  Histograms
show, clockwise from upper-left, the semi-major axis $a$, the
half-mass isophote ellipticity $\epsilon$, the central velocity
dispersion $\sigma$, and the maximum velocity along the major
axis V$_{\rm maj}$.  Each histogram represents the composite properties
of all fifteen remnants, viewed from 150 projections.  The 
semi-major axis histogram includes the half-mass radius of the
progenitor disk.  The ellipticity histogram includes data on
observed ellipticities from \citet{Fab89} and 
\citet{FIZ91}.
\label{fig:allhist}}
\end{figure*}

Figure~\ref{fig:allhist} shows histograms of the semi-major axis $a$,
half-mass isophote ellipticity $\epsilon$, central velocity dispersion
$\sigma$, and maximum rotation speed along the major axis $V_{\rm
maj}$.  Each of the 30 remnants (15 dissipational, and 15 dissipationless)
is projected along 195 lines of sight.  The angles are selected
using the HEALPIX \citep{Gor05} software with $nside=4$ and uniformly 
sample a unit sphere.  This procedure results in a total of
2250 ``data'' points which are displayed in Figure~\ref{fig:allhist}
for both the dissipationless and the dissipational merger remnants,
represented by the ($blue$) open and ($red$) cross-hatched histograms,
respectively.

It is apparent that dissipational merger remnants are quite different
from their dissipationless counterparts.  The effects of gaseous
dissipation and star formation result in remnants that are more
compact, rounder, have higher velocity dispersion, and have a much more
uniform distribution of rotation speeds.

The size of the remnants is a strong function of dissipation.  In
fact, {\it every} dissipational remnant is smaller (by nearly
$\sim50$\%) than its corresponding dissipationless version.  This fact
is unsurprising given the significant gas fraction of our progenitor
disks and the efficiency with which mergers drive gas into the
galaxy centers \citep{BH91,BH96}, where it fuels a burst of star formation
\citep{MH96}.  For the mergers we present here, this process is so
efficient that the average half-mass radius $R_e$ of the dissipational
remnants, 4.0 kpc, is smaller than the original progenitor
disk 4.4 kpc.  The average $R_e$ of the dissipationless
remnants is 6.9 kpc.  The average semi-major axis $a$ of the
dissipational remnants is 4.1 kpc and 7.9 kpc for the
dissipationless remnants.

To check the absolute size of our merger remnants, we note that the
size-mass relationship found in the Sloan Digital Sky Survey (SDSS)
for early-type galaxies suggests a typical size of 3.4 kpc for
galaxies of equivalent stellar mass ($\sim8 \times 10^{10}$\msun)
to our simulated remnants
\citep{Shen03}.  Hence, it appears that both our dissipational and
dissipationless remnants are larger than SDSS galaxies \citep[see
also][]{Rob06fp}.  However, as pointed out by \citet[][Appendix
A]{BK05}, sizes can be systematically underestimated depending on the
range of radii considered when fitting an assumed profile.  We also
note our initial disk galaxies are relatively large, a reflection of
an above average spin parameter $\lambda$ assumed for the dark matter
halo.  In reality there exists a distribution of spins and thus disk
sizes.  Furthermore, \citet{Rob06fp} have shown that the sizes of the
remnants depend somewhat on the orbit of the merger, in agreement
with \citet{BK05}, and matching the observations would thus require a
distribution of orbits consistent with those expected cosmologically.
Therefore, we consider this (dis)agreement as marginal at best and
here we simply note that the dissipational remnants are in closer
agreement with the SDSS size-mass relation for early-type galaxies
than dissipationless remnants.

The top-right plot in Figure~\ref{fig:allhist} shows the distribution
of ellipticities for our merger remnants.  Overplotted are 150
ellipticities as measured and compiled by \citet{FIZ91}, and another
420 elliptical galaxies observed by \citet{Fab89}.
The dissipational merger remnants are nearly indistinguishable from
the two samples of observed ellipticals.  Both the observed and
dissipational ellipticity distributions peak near 0.2 and then drop
precipitously above 0.4.  There exist almost no highly elliptical
($\epsilon>0.6$) galaxies.  These features are in sharp contrast to the
dissipationless remnants whose distribution of ellipticities peaks
near 0.4 and extends beyond 0.7, similar to that found by \citet{NB03}.

The central velocity dispersion is a common measure of galaxy mass and
appears to be correlated with many fundamental properties of
elliptical galaxies.  As the histogram of remnant velocity dispersions
in Figure~\ref{fig:allhist} shows, the central velocity dispersion is
highly dependent upon viewing angle, merger orbit, and the dissipative
effects of gas.  The distribution of dissipationless dispersions is
well fitted by a Gaussian with mean 150~\kms and variance 11~\kms.  The
distribution of dissipational merger remnants is better described as
bi-model, with one component a Gaussian centered at 178~\kms and
variance 10~\kms, and a second component that extends to lower
velocity dispersion.  This second, lower dispersion, distribution is
comprised of five mergers ($d$,$f$,$i$,$l$,$m$) that are all
significant rotators.  The mean and variance of the entire
dissipational distribution is 169~\kms and 15~\kms, respectively.

Because it probes the gravitational potential, the central velocity 
dispersion is often used in combination with the effective radius as 
a proxy for galaxy mass (i.e., $M \propto \sigma^2 R_e$).  While it 
does not affect the results of this paper, we note that $\sigma^2 R_e$
for our remnants can vary by a factor of 2, even though all the remnants
galaxies have exactly the same {\it total} mass.

\begin{figure}
\plotone{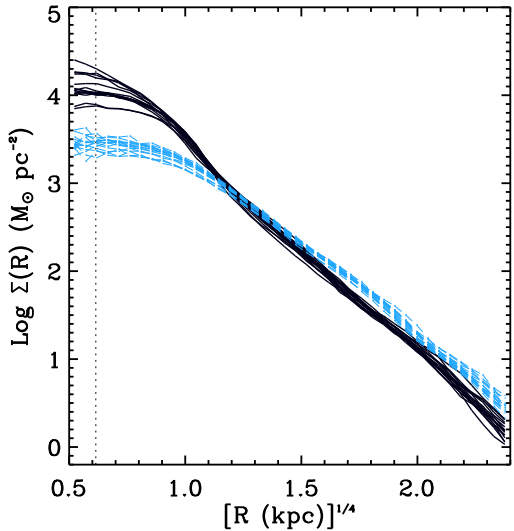}
\caption{The surface density profile for all thirty 
(the fifteen orientations from Table~\ref{tab:orbs}
for both the dissipational and dissipationless
mergers) remnants plotted against $R^{1/4}$ as viewed
from a single projection normal to the orbital plane.
The dark solid lines that have uniformly higher
central surface density are the dissipational merger 
remnants.  The light dashed lines with lower central 
surface density of the dissipationless merger remnants.
The dashed vertical line denotes the softening length,
and hence the resolution of the simulation.
\label{fig:prof}}
\end{figure}

The increased velocity dispersion of the dissipational remnants is
accompanied by a steeper surface density profile, as shown in
Figure~\ref{fig:prof}.  Within a radius of $\sim1.8$~kpc the 
dissipational remnants have a much higher surface density owing to
gas dissipation and star formation.  This feature is present
regardless of the merger orbit, even though the different orbits
produce a variety of central ($<1$~kpc) profiles.  It is intriquing 
that the surface densities appear to be well fit by a $R^{1/4}$-profile and
do not demonstrate the distinct cusp of newly formed stars seen in
previous simulations \citep{MH94dsc,Sp00,Cox05}.  Future work is 
underway to address whether these surface density profiles are
consistent with observed ellipticals and merger remnants
\citep[see, e.g.,][]{RJ04,Lau05} as well as whether the profiles
depend on progenitor disk gas fraction or mass.


Spheroidal (and possibly all) galaxies are thought to host black holes
in their centers in accord with the $M_{\rm BH}-\sigma$ relation
\citep{FM00,Geb00}.  For the remnants displayed here, with central
velocity dispersions $\sim$150~\kms, each remnant should contain a
fairly massive ($\sim$ several $\times 10^7$\msun) black hole.  During
the dissipational mergers, black holes grow to the appropriate mass and
futher accretion is terminated by a feedback induced galactic wind.  Our
merger are thus consistent with the idea that the $M_{\rm BH}-\sigma$ 
relation is produced by the self-regulated growth of the black hole
\citep{dMSH05,SdMH05,Rob06b,Kaz05}.  Because of the lack of gas, the
black holes in the dissipationless mergers remain the same size during
the entire simulation.  In this case, the initial black hole masses
require some fine-tuning in order for the remnants to reside on the
observed $M_{\rm BH}-\sigma$ relation.  We consider this further evidence
for the necessity of gas-rich progenitor disk galaxies.

The final histogram in Figure~\ref{fig:allhist} shows the maximum
rotation velocity along the major axis of each remnant.  As with the
previous histograms, the dissipational and dissipationless remnants
are quite different.  While the dissipationless remnants rotate very
slowly, with all measurements below 50~\kms, the dissipational
remnants span a wider range of rotation speeds, with some projections
reaching speeds of $\sim140$~\kms.  We will discuss the rotational
properties of our merger remnants in more detail in the following
section.

%
\subsection{Rotational Support}
\label{ssec:majrot}

As stated in the introduction, elliptical galaxies can be split
into two groups.  Small ellipticals tend to rotate along their major
axes, while large elliptical galaxies show little or no rotation.
More specifically, the rotation of small ellipticals is consistent
with that of an oblate, isotropic rotator.  Thus, their elliptical
shape can be understood solely in terms of flattening induced by their
observed rotation \citep{Bin78,BT}.  Large ellipticals, on the other
hand, have little rotation.  Their elliptical structure must be
supported by something other than rotation, and is most likely
velocity anisotropy.  One of the clearest ways to visualize this trend
is to plot the rotation velocity $V_{\rm maj}$ divided by the central
velocity dispersion $\sigma$, against the ellipticity $\epsilon$, in
what is commonly referred to as the anisotropy diagram.

\begin{figure*}
\plottwo{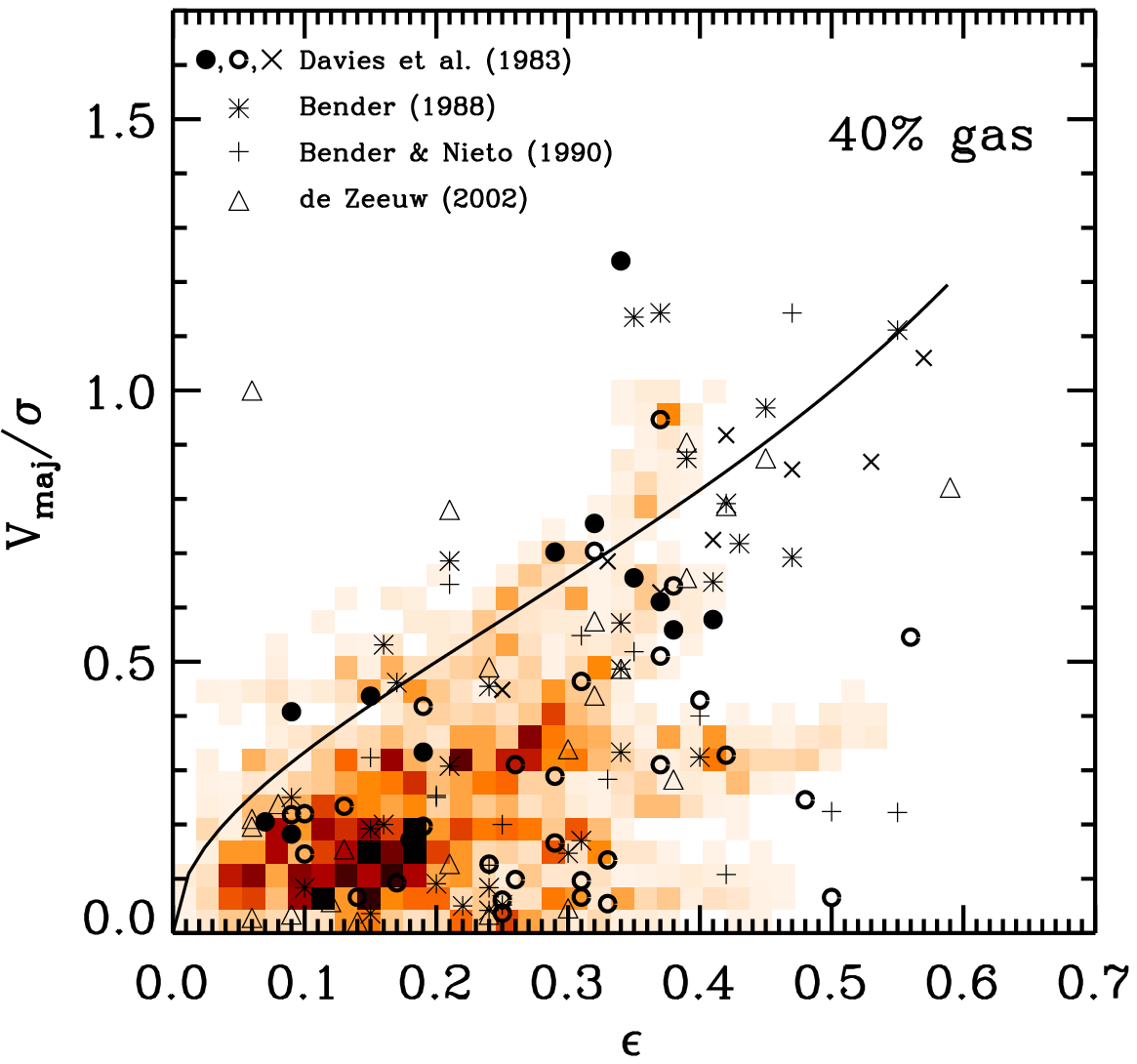}{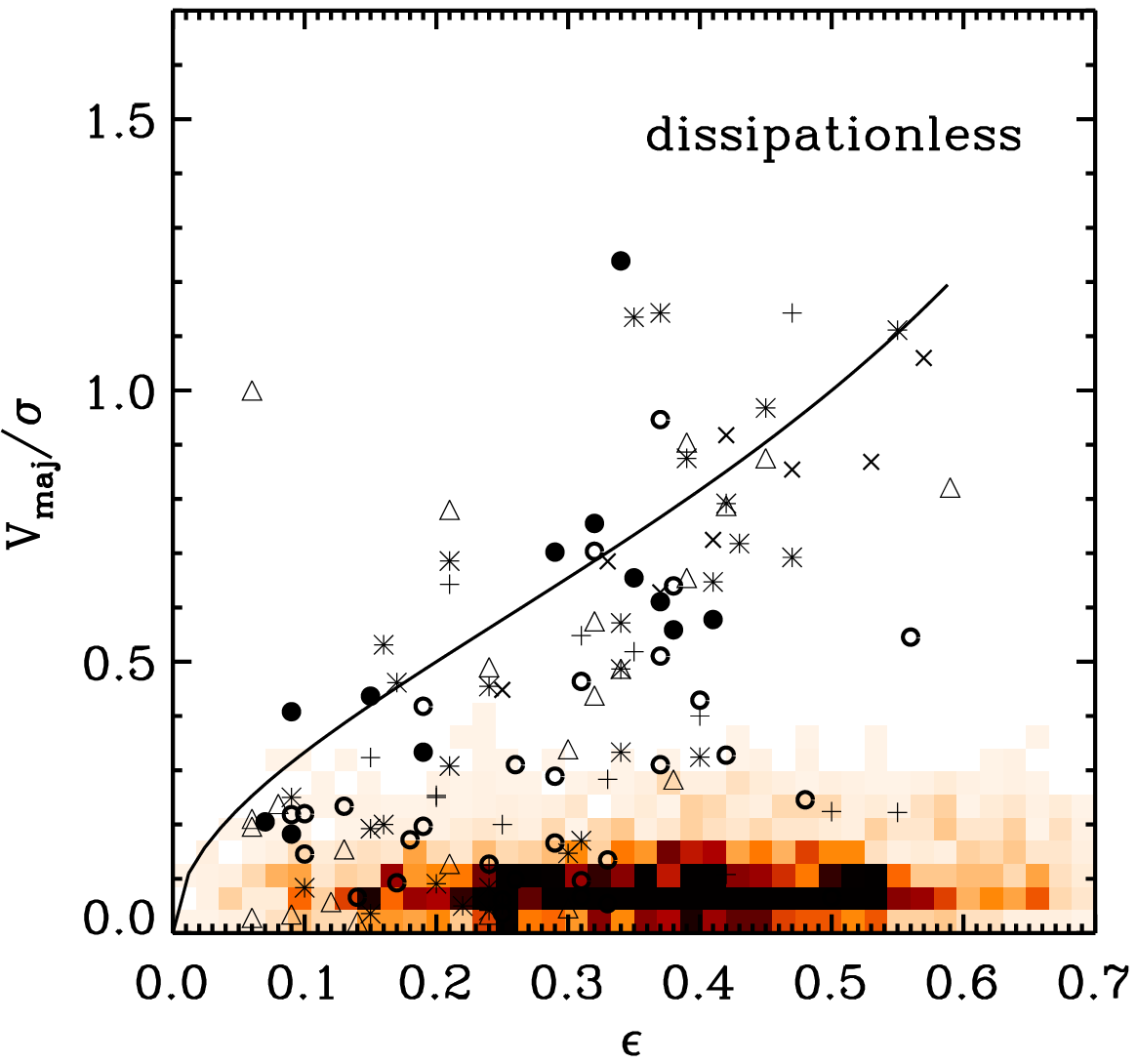}
\caption{The $V_{\rm maj}/\sigma$ versus ellipticity
diagram for dissipational (40\% gas) and dissipationless
merger remnants.  $V_{\rm maj}$ is the maximum rotation
speed measured in a slit along the major axis, $\sigma$
is the velocity dispersion averaged within half of an
half-mass radius, and the ellipticity is measured at the
half-mass isophote.  Further details can be found in 
\S~\ref{ssec:anal}.  The solid line in both plots is
that expected for an oblate isotropic rotator \citep{Bin78}.
Overplotted are data from observed ellipticals from 
\citet{Dav83,Ben88,BN90,dZ02}.
\label{fig:vsig_both}}
\end{figure*}

Figure~\ref{fig:vsig_both} shows our merger remnants in an anisotropy
diagram.  For reference, overplotted on this diagram are data from
several observational samples \citep{Dav83,Ben88,BN90,dZ02}.
\citet{Bin78} has shown that for oblate isotropic rotators $V/\sigma$
is purely a function of ellipticity $\epsilon$.  If this is the case,
then at high ellipticities one is viewing a rotating, flattened
isotropic ellipsoid from edge on.  The solid line in
Figure~\ref{fig:vsig_both} is this relation for oblate isotropic rotators
\begin{equation}
(V/\sigma) =
[\epsilon/(1-\epsilon)]^{1/2}. \label{eq:vsigoir} 
\end{equation}
\citep{Bin78}.  In terms of the anisotropy diagram, galaxies that are
near the line are presumably oblate isotropic rotators.  Their shapes
are flattened by rotation and they are being viewed from a variety of
angles with respect to their rotation axes.  Galaxies that are below
this line, i.e. those that have very little rotation, and some amount
of ellipticity, are likely to have strong velocity anisotropies.

As in the previous section, the dissipationless and dissipational
remnants occupy very different regions in the anisotropy diagram.  The
dissipationless remnants, presented in the right-hand plot of
Figure~\ref{fig:vsig_both}, occupy a nearly identical region of the
anisotropy diagram as those presented in
\citet{NB03}.  Consistent with their findings, most remnants are very
elliptical and hardly rotating \citep[see also][]{NW83,B88,HRemI}.  
We note that our additional orbits
($b-h$) produce non-rotating remnants exactly like orbits $i-p$ (i.e.,
the exact orbits simulated by \citet{NB03}).  There are a large number of
remnants that have an ellipticity greater than 0.6, as was noted by
\citet{NB03}, and was clearly displayed in the histogram of
Figure~\ref{fig:allhist}.  This distribution of ellipticities is
not observed in real ellipticals, especially for galaxies with
little or no rotation.  Ellipticity aside, the small rotation of the
dissipationless merger remnants led \citet{NB03} to suggest dissipationless
mergers as a possible formation mechanism for bright ellipticals.

The dissipational merger remnants are shown in the anisotropy diagram on
the left panel of Figure~\ref{fig:vsig_both}.  These remnants span a
much larger range of rotational properties than the dissipational
remnants, and include remnants that reside on the oblate isotropic
rotator line as well as remnants that fall below this line.  The most
common position for dissipational remnants to reside in the anisotropy
diagram is $V/\sigma\sim0.35$ and $\epsilon\sim0.18$.  At this point
the remnant is slowly rotating, is below the oblate isotropic rotator
line, and its structure is presumably supported by anisotropy; all of
these are features similar to bright ellipticals.

\begin{figure}
\plotone{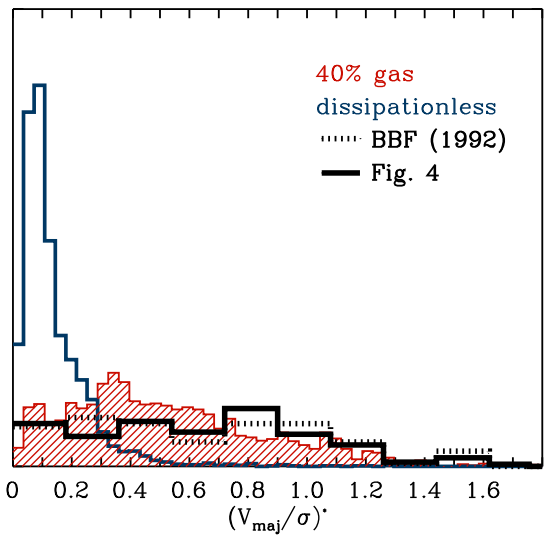}
\caption{Histogram of (V$_{\rm maj}$/$\sigma$)$^*$ for both
the dissipationless (open, $blue$) and dissipational (filled, $red$) merger 
remnants.  Also shown is a histogram of observed ellipticals
for the same points shown in Figure~\ref{fig:vsig_both} and a 
histogram of the sample of spheroids compiled by \citet{BBF92}.
\label{fig:vsig_hist}}
\end{figure}

In general, Figure~\ref{fig:vsig_both} shows that the
dissipational remnants encompass a significant fraction of the
observational data points; much more so than the dissipationless
remnants.  To investigate this further, for each one of our merger
remnants we evaluate ($V_{\rm maj}/\sigma$)$^*$, the ratio of the
measured rotation parameter $V_{\rm maj}/\sigma$ to
Equation~(\ref{eq:vsigoir}), the value for an isotropic oblate spheroid
flattened by rotation \citep{Bin78,Dav83,KB96}.  In a manner similar
to Figure~\ref{fig:allhist} we show histograms of ($V_{\rm
maj}/\sigma$)$^*$ for both the dissipationless and dissipational
merger remnants in Figure~\ref{fig:vsig_hist}.  In addition we show
the distribution for the same observations as shown in
Figure~\ref{fig:vsig_both} as well as the observational sample of
\citet{BBF92}.  We caution that these samples are not independent of
one another as there are a number of galaxies which are counted in
both.  Confirming the trend in Figure~\ref{fig:vsig_both}, 
we see that the observations and
dissipational remnants have a predominantly flat distribution below
($V_{\rm maj}/\sigma$)$^* < 1$, and a tail of values above this.  In
contrast, the dissipational remnants are peaked below 0.2 and there
are almost no remnants above 0.6.

Even though the dissipational remnants are a much better match to the
observed distribution of ($V_{\rm maj}/\sigma$)$^*$, as shown in
Figure~\ref{fig:vsig_hist}, there is still a discrepancy between the
observed and predicted number of fast rotators; i.e., 
those with ($V_{\rm maj}/\sigma$)$^*
\geq1$.  This deficiency was also noted by several other studies under
a variety of different circumstances \citep{Cre01,GG05a,GG05c,NB03},
even when minor mergers were included.  Observations find a large
number of galaxies exhibiting fast rotation.  In the observations
plotted here, 20\% of the galaxies shown in Figure~\ref{fig:vsig_both}
have ($V_{\rm maj}/\sigma$)$^*>1$, and this number increases to 33\%
for the \citet{BBF92} sample which includes a larger number of faint
ellipticals and bulges.  The fraction of fast rotators could increase
even more with the inclusion of S0s and observations which probe the
velocity field at large radii \citep{RCF99}.  For our simulated
dissipational merger remnants, only 11\% of the projected images have
($V_{\rm maj}/\sigma$)$^*>1$.  For the moment, we do not find this
discrepancy alarming as we have sampled a limited range of disk galaxy
initial conditions.  In particular, we suspect that varying the gas
fraction may produce a much larger fraction of fast rotating remnants,
continuing the trend seen in Figure~\ref{fig:vsig_both}.  We briefly
address this in \S~\ref{ssec:dep}.

%
\subsection{The Origin of Stellar Rotation}
\label{ssec:origin}

%
%
\begin{figure}
\plotone{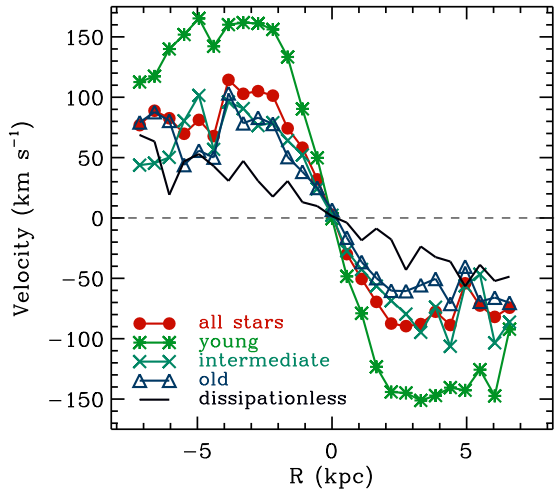}
\caption{The rotation curve for the dissipational merger remnant $f$,
viewed perpendicular to the orbital plane.  This is one of the
fastest rotating remnants and is identical to that used in 
Figures~\ref{fig:remimg} and \ref{fig:f_time}.  The rotation
curve is decomposed by particle type according to the key inset
within the bottom-left of the plot.  The dissipational stellar 
particles are segregated based upon their relative age;
``old'' stars begin the simulation as a collisionless particle,
``intermediate'' stars were created during the isolated and
first passage phases, 
while ``young'' stars are created from dense gas during the
final merger.  For reference the composite ``all stars'' shows
the total stellar rotation curve, and the solid black line
shows the dissipationless merger remnant viewed from the same
angle with an identical slit.
\label{fig:slitvel}}
\end{figure}

To understand what generates the significant rotation found in
dissipational remnants, we decompose the stellar rotation curve
presented in the lower-left of Figure~\ref{fig:remimg} into
contributions from stars by their age.  The resulting rotation curves
are shown in Figure~\ref{fig:slitvel}.  Star particles are designated
as ``old'', ``intermediate'', or ``young'' on the basis of when they
originally became a dissipationless stellar particle.  Old stars are
those that are present prior to the start of the simulations, and thus
designated as stars when setting up the initial conditions.  Star
formation during the course of the merger simulation predominantly
occurs in bursts that accompany the first passage of the galaxies, and
the final merger event \citep[see e.g.,][]{MH94majm,Cox05,SdMH05}.  These burst
events occur $\sim$1~Gyr apart.  We denote any stars created during
the isolated evolution prior to the first burst, and those generated
during the first burst as ``intermediate''.  All stars created during
the final merger are ``young''.  In addition to the rotation curves for
stars in the dissipational run, we also plot the stars from the
dissipationless simulation as a solid black line in
Figure~\ref{fig:slitvel}.

Figure~\ref{fig:slitvel} demonstrates that it is the stars born during
the final merger, i.e., the ``young'' stars, that are the fastest
rotators.  Because this young stellar component is centrally concentrated
(its half-mass radius is less than 1~kpc) and is the dominant stellar
component inside of $\sim$1~kpc, the central rotation closely tracks
that of the young stars.  Outside of 1~kpc the ``intermediate'' and
``old'' populations dominate the stellar mass and thus the rotation
curve closely tracks these components.  Both ``intermediate'' and
``old'' components have similar rotation suggesting that all stars
that exist prior to the final merger are equivalent in terms of their
rotation properties in the merger remnant.

Figure~\ref{fig:slitvel} also plots the rotation curve for the
dissipationless version of this merger, shown with a solid black line.
While the dissipationless remnant also exhibits rotation, the rotation
curve is much shallower than the dissipational remnant.  Obviously,
gaseous dissipation and star formation have affected the rotation of
the old stellar population in the dissipational remnant.  The most
plausible explanation for this is simply the conservation of angular
momentum.  The central concentration of young stars steepens the
potential well and draws the old stars into the center.  The half-mass
radius of the ``old'' stars is $\sim$20\% less than in the
dissipational remnant and thus the velocity of these stars must be
roughly $\sim$20\% higher to conserve angular momentum.  The fact that
the velocity increase is larger than this may represent a transfer of
angular momentum between the stars and gas, as was shown by
\citet{MH96}.

Even though the young stars rotate faster than the old stars, most of
the angular momentum is carried by the old stars.  Specifically, the
old stars typically have three times the specific angular momentum of
the young stars.  However, most of the angular momentum carried by the
old stars is at radii larger than the effective radius, and thus is
not contained in our slit which only extends slightly beyond one
effective radius.  Figure~\ref{fig:slitvel} shows that beyond
$\sim$4~kpc, the dissipational and dissipationless remnants have
similar rotation, and also similar angular momentum distributions for
the old stellar component.  Thus, these results suggest that the
significant rotation seen in dissipational remnants is not indicative
of more angular momentum, but dissipation and star formation
effectively redistribute angular momentum from large radii to within
the effective radius where rotation is typically measured.

Although our results indicate a segregation between stellar age and
rotation it is unclear if this trend could be verified by
observations.  As the stellar populations evolve, the small age
differences (1-2~Gyr) will quickly be washed out and thus difficult to
disentangle.  However, the general trend for young stars to be centrally
concentrated and kinematically different than the more extended older
stars is a generic feature of these simulations.  In a few cases the
merger results in counter-rotating or kinematically decoupled cores, as
is seen in many observed ellipticals \citep{DB88,FIH89}, and as found
earlier by \citet{HB91} using simulations with a smaller
gas fraction than here and ignoring star formation.

%
\subsection{Individual Merger Remnants}
\label{ssec:ind}

\begin{center}
\begin{deluxetable*}{c|cccrcc|cccrcc}
\tabletypesize{\scriptsize}
\tablecaption{Projection Averaged Remnant Properties\label{tab:remprops}}
\tablewidth{0pt}
\tablecolumns{13}
\tablehead{
\colhead{Orientation} &
\multicolumn{6}{c}{Dissipationless} &
\multicolumn{6}{c}{40\% Gas}\\
\colhead{ID} &
\colhead{$\epsilon$} &
\colhead{R$_e$} &
\colhead{$a$} &
\colhead{$V_{\rm maj}$} &
\colhead{$\sigma$} &
\colhead{$(V_{\rm maj}/\sigma)^*$} &
\colhead{$\epsilon$} &
\colhead{R$_e$} &
\colhead{$a$} &
\colhead{$V_{\rm maj}$} &
\colhead{$\sigma$} &
\colhead{$(V_{\rm maj}/\sigma)^*$}\\
\colhead{} & \colhead{} & \colhead{(kpc)} & \colhead{(kpc)} &
\colhead{(\kms)} & \colhead{(\kms)} & \colhead{} &
\colhead{} & \colhead{(kpc)} & \colhead{(kpc)} &
\colhead{(\kms)} & \colhead{(\kms)} & \colhead{}\\
\colhead{} & \colhead{(1)} & \colhead{(2)} & \colhead{(3)} &
\colhead{(4)} & \colhead{(5)} & \colhead{(6)} &
\colhead{(1)} & \colhead{(2)} & \colhead{(3)} & 
\colhead{(4)} & \colhead{(5)} & \colhead{(6)}
}
\startdata
h & 0.48 &  6.49 & 8.73 & 14.3 &  150.8 & 0.09  & 0.27 &  3.56 & 4.16 &  46.0 &  180.7 & 0.25  \\
b & 0.46 &  6.27 & 8.21 & 11.7 &  150.0 & 0.08  & 0.27 &  3.70 & 4.21 &  26.8 &  182.2 & 0.15  \\
c & 0.50 &  6.37 & 8.39 & 15.8 &  153.5 & 0.10  & 0.32 &  3.70 & 4.43 & 116.6 &  177.5 & 0.65  \\
d & 0.42 &  6.40 & 8.04 &  9.1 &  151.0 & 0.06  & 0.24 &  3.86 & 4.34 & 111.9 &  147.7 & 0.76  \\
e & 0.37 &  6.23 & 7.87 &  9.7 &  149.7 & 0.06  & 0.21 &  3.49 & 3.99 &  11.3 &  178.5 & 0.06  \\
f & 0.33 &  6.16 & 7.17 & 15.3 &  147.0 & 0.10  & 0.26 &  3.99 & 4.54 & 143.3 &  140.5 & 1.03  \\
g & 0.37 &  6.19 & 7.60 & 13.0 &  150.2 & 0.09  & 0.19 &  3.84 & 4.19 &  65.5 &  172.7 & 0.37  \\
i & 0.34 &  6.24 & 7.71 & 13.8 &  150.4 & 0.09  & 0.19 &  3.64 & 4.00 & 115.7 &  155.4 & 0.75  \\
j & 0.22 &  6.09 & 6.91 & 15.5 &  149.6 & 0.10  & 0.13 &  3.36 & 3.63 &  66.0 &  169.6 & 0.39  \\
k & 0.49 &  6.39 & 8.83 &  9.3 &  151.4 & 0.06  & 0.22 &  3.56 & 4.01 &  73.2 &  175.2 & 0.41  \\
l & 0.30 &  6.24 & 7.49 & 27.2 &  147.8 & 0.18  & 0.19 &  4.12 & 4.49 &  91.5 &  147.8 & 0.62  \\
m & 0.39 &  6.21 & 7.99 & 14.0 &  151.4 & 0.09  & 0.17 &  4.01 & 4.30 &  77.2 &  146.2 & 0.53  \\
n & 0.26 &  6.29 & 7.34 & 12.3 &  150.4 & 0.08  & 0.21 &  3.50 & 3.93 & 131.4 &  169.4 & 0.77  \\
o & 0.35 &  6.20 & 7.46 & 31.2 &  146.6 & 0.21  & 0.20 &  3.67 & 4.11 & 128.7 &  166.4 & 0.77  \\
p & 0.45 &  6.29 & 8.21 &  8.0 &  152.1 & 0.05  & 0.24 &  3.59 & 4.07 &  66.0 &  168.6 & 0.39  \\
\enddata
\tablecomments{  
List of properties of simulated merger remnants averaged from 190
viewing angles selected to uniformly sample the unit sphere.  Column (1) is the 
ellipticity of the half-mass isophote.  Column (2) is the half-mass radius as 
measured using circular apertures.  Column (3) is semi-major axis of the half-mass
isophote.  Column (4) is the maximum rotation speed along the major axis. Column
(5) is the central velocity dispersion, averaged within 0.5$a$, and column (6)
is the ratio of $V_{\rm maj}/\sigma$ to that expected from an oblate isotropic
rotator.
}
\end{deluxetable*}
\end{center}

\begin{figure*}
\plotone{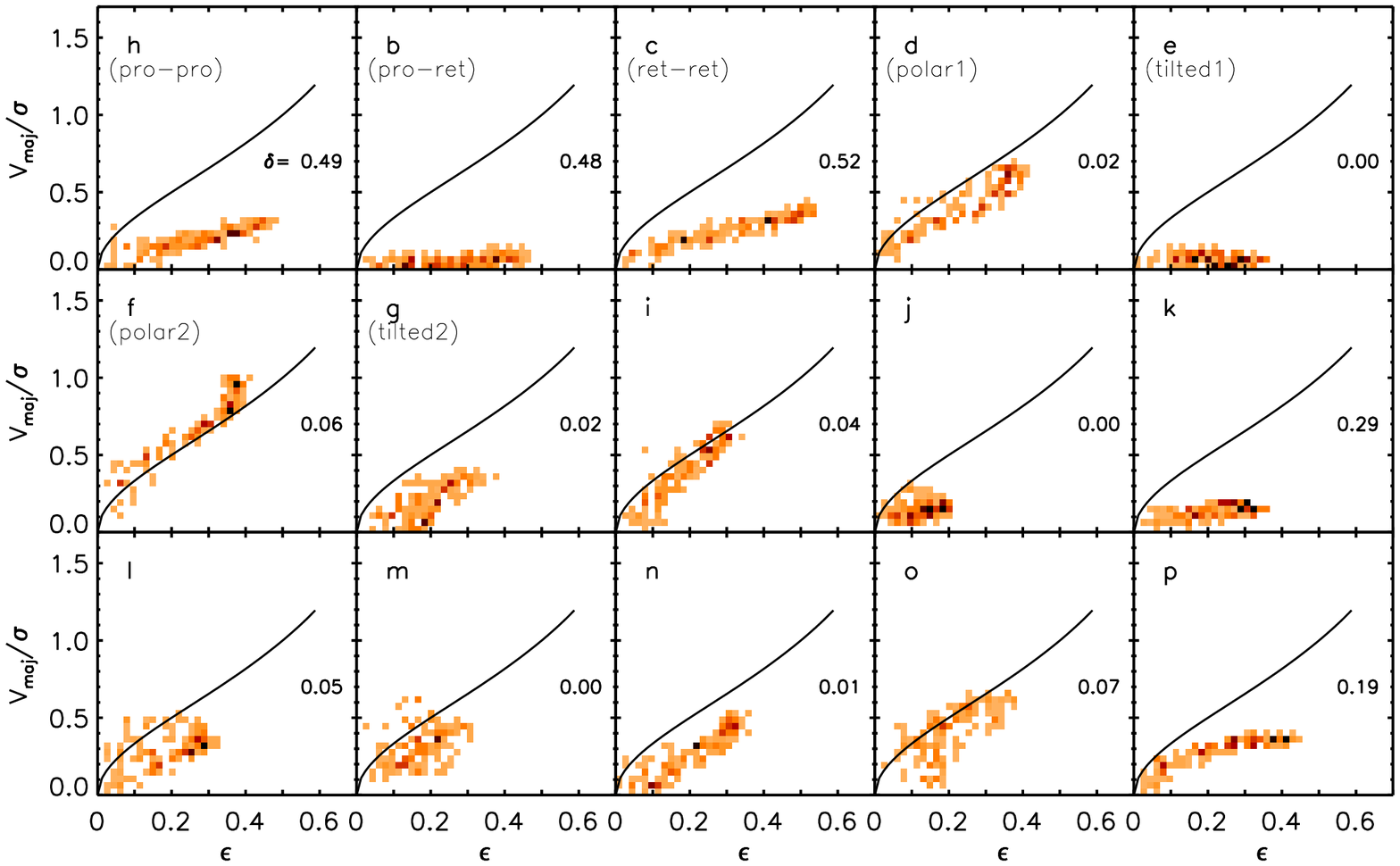}
\plotone{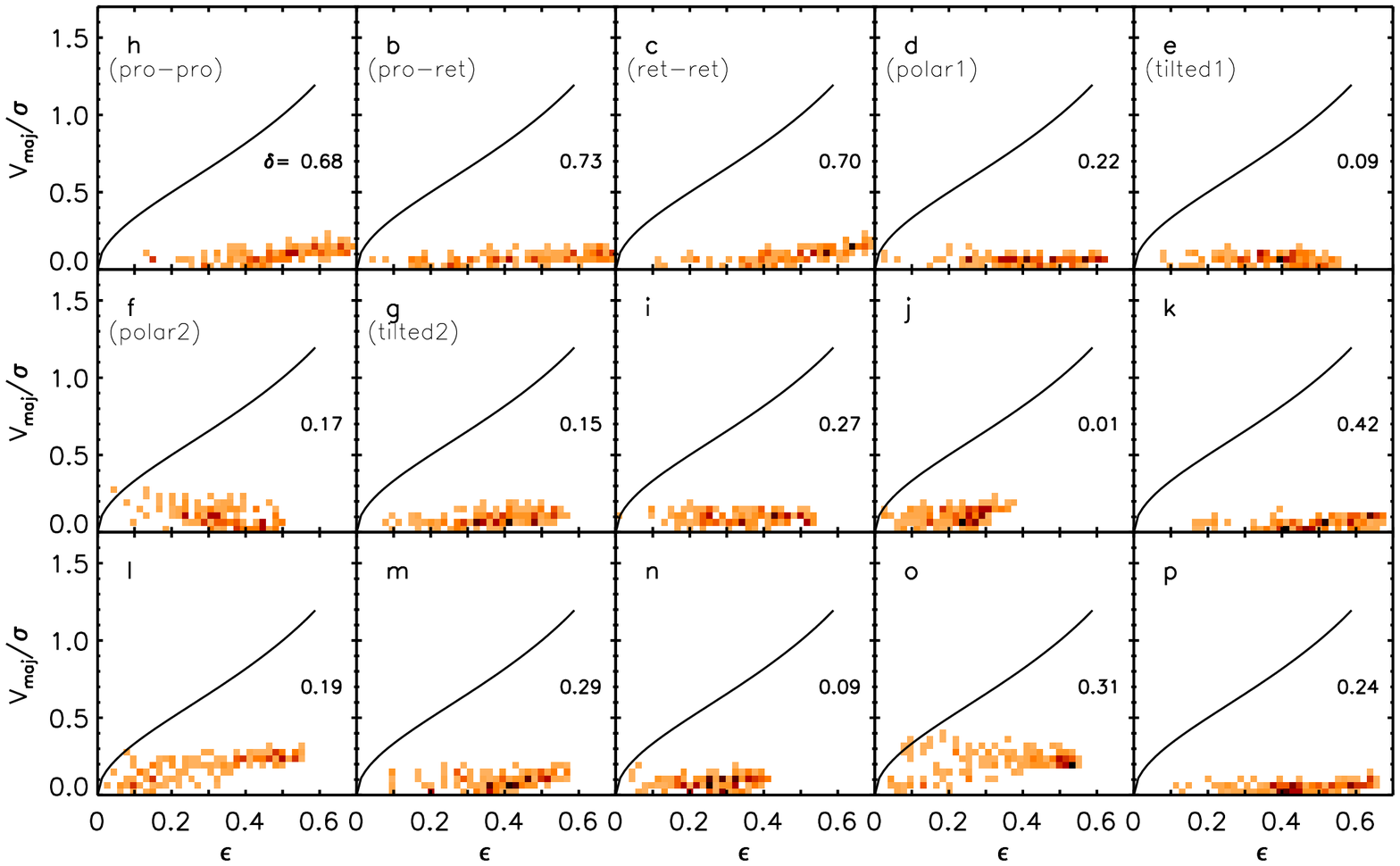}
\caption{$V_{\rm maj}/\sigma$ versus ellipticity, or anisotropy,
diagram for individual merger remnants.  The top grid of fifteen
panels shows remnants where the progenitor disk contained
40\% gas and the bottom grid is the corresponding mergers
when the progenitor disks are dissipationless.  Each 
plot shows one merger remnant for one of the merger
orientations listed in Table~\ref{tab:orbs}, viewed from 190
different viewing angles chosen to uniformly sample the unit
sphere.  The solid line is that expected for an oblate isotropic
rotator and the number in the middle-right of each panel is
the anisotropy parameter $\delta$ defined in 
\S~\ref{ssec:anisot}.  For reference, $\delta=0$ is isotropic
and $\delta>0$ is a sign of anisotropy.
\label{fig:vsig_orb}}
\end{figure*}

Although dissipational remnants display much more rotation, on
average, than their dissipationless counterparts, there is a large
spread of rotation properties between individual remnants
\citep{BB00}.  Figure~\ref{fig:vsig_orb} demonstrates this behavior by
displaying the rotational properties of individual merger remnants,
one per panel, viewed from 190 projections.  It is these remnants that
are co-added to produce Figure~\ref{fig:vsig_both}.

Figure~\ref{fig:vsig_orb} clearly shows that projection effects cause
each remnant to occupy a band in the anisotropy diagram.  Every
remnant appears nearly circular from a small range of viewing angles
and elliptical from most viewing angles.  It is typically the case
that remnants rotate very little when they appear circular and thus
each band passes near the origin of the anisotropy diagram.  At larger
projected ellipticities there exists a wide range of rotational
properties for the fifteen merger remnants.  In some cases ($d$, $f$,
$i$, and $o$) the remnant closely resembles an oblate isotropic
rotator.  However, in most cases the remnant swath is below the oblate
isotropic rotator line.  In all cases (except, perhaps, the $j$ merger
remnant) the maximum rotation is coincident with the largest projected
ellipticity, suggesting that the majority of the dissipational
merger remnants are (nearly) oblate and the rotation axis is along
their short axis.

It is important to emphasize the large range of kinematic properties
present in the dissipational merger remnants.  This range of remnant
properties owes entirely to the variations in initial disk
orientations.  Specifically, it appears that the co-planar orbits
($h$, $b$, and $c$) produce very flattened anisotropic remnants, while
polar orbits, i.e., where one disk orbits in a 90$^\circ$ orientation
to another disk ($d$ and $f$), result in fast rotating remnants that
are oblate and isotropic.  Finally, tilted orbits, such as $e$, $j$,
and $k$ produce fairly round remnants that are not rotating.

The dissipationless merger remnants shown in Figure~\ref{fig:vsig_orb}
show much less variability than the dissipational remnants.  In all
cases, the remnants span a wide range of ellipticities and are never
rotating.  Elliptical remnants that do not rotate are thought to owe
their ellipsoidal shape to anisotropic velocity dispersions.  We will
investigate this in the following section.

A final point concerns the initial disk orientations and whether or
not they faithfully represent the full range of possibilities.
Orientations $i-p$ were originally selected to be a uniform sampling
of disk orientations \citep[see][]{B92}.  However, as we note above,
the co-planar orbits ($h$, $b$, and $c$) and polar orbits ($d$ and
$f$) produce somewhat peculiar remnants that are not well represented
by the uniform sampling introduced by Barnes.  Hence, it seems that
one must include both a uniform grid to represents the
majority of merger events as well as a small sample of hand-selected
orientations to represent more pathological orbits in order to gain 
a full understanding of the various properties
possible in the aftermath of a galaxy collision.

%

\begin{figure}
\plotone{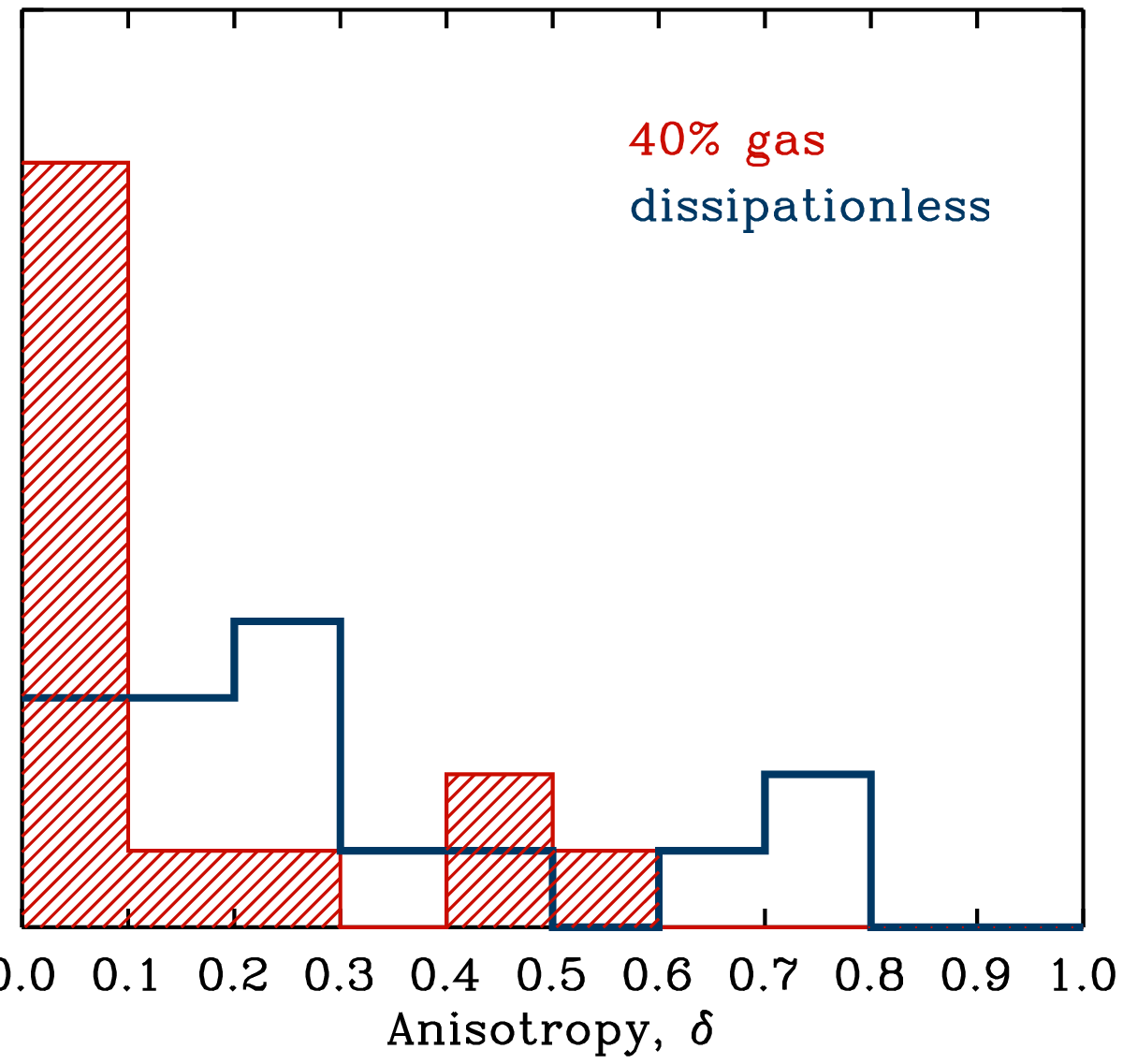}
\caption{Histogram of remnant velocity anisotropy parameters $\delta$,
where $\delta$ is calculated via Equation~(\ref{eq:delta}).  With this
definition for anisotropy, $\delta=0$ is isotropic and $\delta>0$
is anisotropic. 
\label{fig:delta}}
\end{figure}

\subsection{Anisotropy}
\label{ssec:anisot}

In the previous section, we demonstrated that all of the
dissipationless and several of the dissipational merger remnants are
ellipsoidal and slowly rotating.  In short, they reside below the
oblate isotropic rotator line in the anisotropy diagram.  In this case,
it is assumed that the ellipsoidal shapes are supported by an
anisotropic velocity dispersion.  However, it has recently been shown
that mergers between unequal mass galaxies can produce fast rotators
that are very anisotropic \citep{BN05}, and thus we wonder if the
slowly rotating remnants are necessarily anisotropic.  In this section
we attempt to address this question by measuring the velocity
anisotropy of each merger remnant.

We note that any dissipationless system in equilibrium can be
accurately described by the tensor virial theorem
\begin{equation}
2T_{jk} + \Pi_{jk} = -W_{jk},
\label{eq:tvt}
\end{equation}
where $W_{jk}$ is the potential energy tensor, and $T_{jk}$ and
$\Pi_{jk}$ are the ordered and random components, respectively, of the
kinetic energy tensor \citep[eq. 4-78]{BT}.  Taking the trace of
Equation~(\ref{eq:tvt}) results in the scalar virial theorem, which
simply relates the total kinetic energy to the total potential energy.

To calculate the kinetic energy tensor we assume each N-body stellar
particle has a gaussian velocity distribution.  The mean and distribution
are computed by kernel weighting among the 96 nearest neighbors.
A Cartesian coordinate system is selected such that the z-axis is 
parallel to the original orbital angular momentum and the x and y
axes are arbitrary.  The mean velocity then contributes to $T_{jk}$
and the dispersion to $\Pi_{jk}$.  For our merger remnants the
kinetic energy is predominantly in the form of random motions, with
\begin{equation}
\frac{{\rm Trace}(\Pi_{jk})}{{\rm Trace}(T_{jk})} \geq3.
\label{eq:tk}
\end{equation}
In comparison, this ratio for the progenitor disks is $\sim0.17$.

The anisotropy of each remnant, then, is the degree to which the
elements of $\Pi_{jk}$, the velocity dispersion tensor, are unequal.
Owing to our initial orbital configuration, in which the orbital
angular momentum is aligned with the z-axis, we typically find
$\Pi_{xx},\Pi_{yy} > \Pi_{zz}$.  We follow \citet{BN05} and define the
anisotropy of the system as
\begin{equation} \delta = 
1-\frac{2\Pi_{\rm zz}}{\Pi_{\rm xx} + \Pi_{\rm yy}}.
\label{eq:delta}
\end{equation}
With this definition, an isotropic system has $\delta=0$, and
$\delta>0$ indicates some degree of anisotropy.  For two of the
dissipational remnants ($g$ and $n$), $\delta$ is slightly negative.
However, these remnants are nearly isotropic and in this case we quote
the absolute value of $\delta$.

A histogram of remnant anisotropies is presented in
Figure~\ref{fig:delta} and the anisotropy of individual merger
remnants is printed in each panel of Figure~\ref{fig:vsig_orb}, along
with the swath of $V_{\rm maj}/\sigma$ owing to various projections.
From Figure~\ref{fig:delta} we see a systematic decrease in $\delta$ for
dissipational remnants compared to dissipationless remnants.  In other
words, every dissipational remnant is much closer to isotropic than
its dissipationless counterpart.  Roughly two-thirds (9/15) of the
dissipational remnants are consistent with being isotropic
($\delta<0.05$) compared to only one of the dissipationless remnants
($j$).  Although there is a fair amount of scatter, there seems to be
a uniform change in the anisotropy for dissipational mergers
equivalent to $\Delta\delta\sim-0.16$ compared to dissipationless
remnants.

As with the rotation, Figure~\ref{fig:vsig_orb} also shows that the
initial disk orientations have a large effect on the remnant
anisotropy.  This fact is perhaps not that surprising given the strong
link between rotation and anisotropy.  In particular, the three
initial orientations in which the initial disks are aligned with the
orbital plane, i.e., co-planar mergers, ($h$, $b$, and $c$) result in
remnants that are much more anisotropic than any of the other
orientations.  This trend appears to hold regardless of the presence
of gas or not.

Figure~\ref{fig:vsig_orb} also allows us to assess the relation
between our (rough) measure of anisotropy $\delta$ and rotation, as
measured by the anisotropy diagram.  As outlined in
\S~\ref{ssec:majrot} these quantities should be anti-correlated,
so that fast rotators are likely to be isotropic, while flattened
non-rotators should be anisotropic.  However, this relation has not
been tested in numerical experiments such as ours.

At first glance, we note that the three dissipational remnants ($d$,
$f$, and $i$) whose projected rotation is consistent with the line for
oblate isotropic rotators (the solid line in the anisotropy diagram)
are also isotropic ($\delta<0.05)$.  Next, many of the remnants that
show very little projected rotation (dissipational remnants $h$, $b$,
$c$, $k$, $p$ and every dissipationless remnant except $j$) are all
anisotropic.  Therefore, for at least 22 of the 30 total merger
remnants, the expected relation between anisotropy and rotation
appears to hold.

However, there remain eight remnants for which their rotation
properties are difficult to reconcile with their anisotropy.  Of these
eight, only one dissipationless remnant seems out of place; $j$.  This
merger remnant is not rotating and is moderately flattened (average
ellipticity of 0.22) yet is apparently isotropic.  One of the
dissipational remnants, $e$, is similar.  The six remaining
outlyers are all dissipational remnants ($g$, $j$, $l$, $m$, $n$,
and $o$) and are all quite similar.  These remnants are round (average
ellipticity $\approx0.18$), and have a large range of rotational
properties for any one projected ellipticity.  As such these remnants
appear to occupy more extended regions within the anisotropy diagram,
rather than distinct swathes.  Future work will be necessary to determine
whether the measure of anisotropy provided by equation~(\ref{eq:delta}) 
is sufficient to describe these systems as well as to determine what
generates their anomalous properties.

%

\begin{deluxetable}{l|rrr|rrr}
\tabletypesize{\scriptsize}
\tablecaption{Remnant Shapes\label{tab:shapes}}
\tablewidth{0pt}
\tablehead{
\multicolumn{4}{r}{Dissipationless} &
\multicolumn{3}{r}{40\% Gas}\\
\colhead{ID} &
\colhead{$b$} & \colhead{$c$} & \colhead{$T$} &
\colhead{$b$} & \colhead{$c$} & \colhead{$T$}
}
\startdata
h & 0.73 & 0.42 & 0.56 & 0.94 & 0.43 & 0.14  \\
b & 0.86 & 0.39 & 0.30 & 0.94 & 0.42 & 0.14  \\
c & 0.75 & 0.37 & 0.50 & 0.98 & 0.40 & 0.04  \\
d & 0.77 & 0.55 & 0.58 & 0.90 & 0.67 & 0.34  \\
e & 0.76 & 0.62 & 0.68 & 0.85 & 0.71 & 0.57  \\
f & 0.78 & 0.67 & 0.67 & 0.94 & 0.40 & 0.41  \\
g & 0.79 & 0.62 & 0.61 & 0.86 & 0.74 & 0.57  \\
i & 0.82 & 0.60 & 0.52 & 0.93 & 0.69 & 0.24  \\
j & 0.84 & 0.76 & 0.73 & 0.96 & 0.82 & 0.24  \\
k & 0.69 & 0.49 & 0.69 & 0.93 & 0.62 & 0.21  \\
l & 0.91 & 0.61 & 0.27 & 0.93 & 0.71 & 0.28  \\
m & 0.72 & 0.60 & 0.75 & 0.88 & 0.71 & 0.44  \\
n & 0.81 & 0.73 & 0.74 & 0.97 & 0.81 & 0.15  \\
o & 0.96 & 0.62 & 0.12 & 0.96 & 0.79 & 0.19  \\
p & 0.76 & 0.52 & 0.58 & 0.93 & 0.63 & 0.23  \\
\enddata
\tablecomments{List of remnant axial ratios
$b$ and $c$, and triaxiality parameter $T$,
as defined in \S~\ref{ssec:shapes}.
}
\end{deluxetable}

\subsection{Shapes}
\label{ssec:shapes}

The results presented in \S~\ref{ssec:basics}, specifically the
distribution of projected ellipticities in Figure~\ref{fig:allhist},
show that, in projection, dissipational remnants tend to be rounder
than dissipationless remnants.  To further quantify this trend we
determine the three-dimensional shape of each remnant as measured by
the axis ratios of the inertia tensor \citep{B92, HRemI,HRemII,Sp00}.
Because our analysis has focused upon the stellar component, primarily
within $R_e$, we use only the most bound half of all stellar particles
to calculate the inertia tensor.  We note that there are many other
methods by which the shapes can be quantified.  Most of these methods
calculate the eigenvectors from some form of the inertia tensor.  In
our case, we calculated the inertia tensor based upon the most bound
stellar particles but alternatives are spherical or ellipsoidal
windows which may be interatively adjusted until the measured shape
and selected aperture agree \citep[see e.g.,][for survey of different
methods and recent results from the literature]{All06}.

The axis ratios $b$ and $c$ are defined as
$b=(\lambda_2/\lambda_1)^{1/2}$ and $c=(\lambda_3/\lambda_1)^{1/2}$,
where $\lambda_1\geq\lambda_2\geq\lambda_3$ are the eigenvalues of the
inertia tensor.  Values of $b$, $c$, along with the ``triaxiality''
parameter $T=(1-b^2)/(1-c^2)$ \citep{FIZ91} are listed in
Table~\ref{tab:shapes} for all of the dissipationless and
dissipational remnants.  For reference, $T=1$ is prolate, and $T=0$ is
oblate, and all values in between are triaxial.

\begin{figure*}
\plottwo{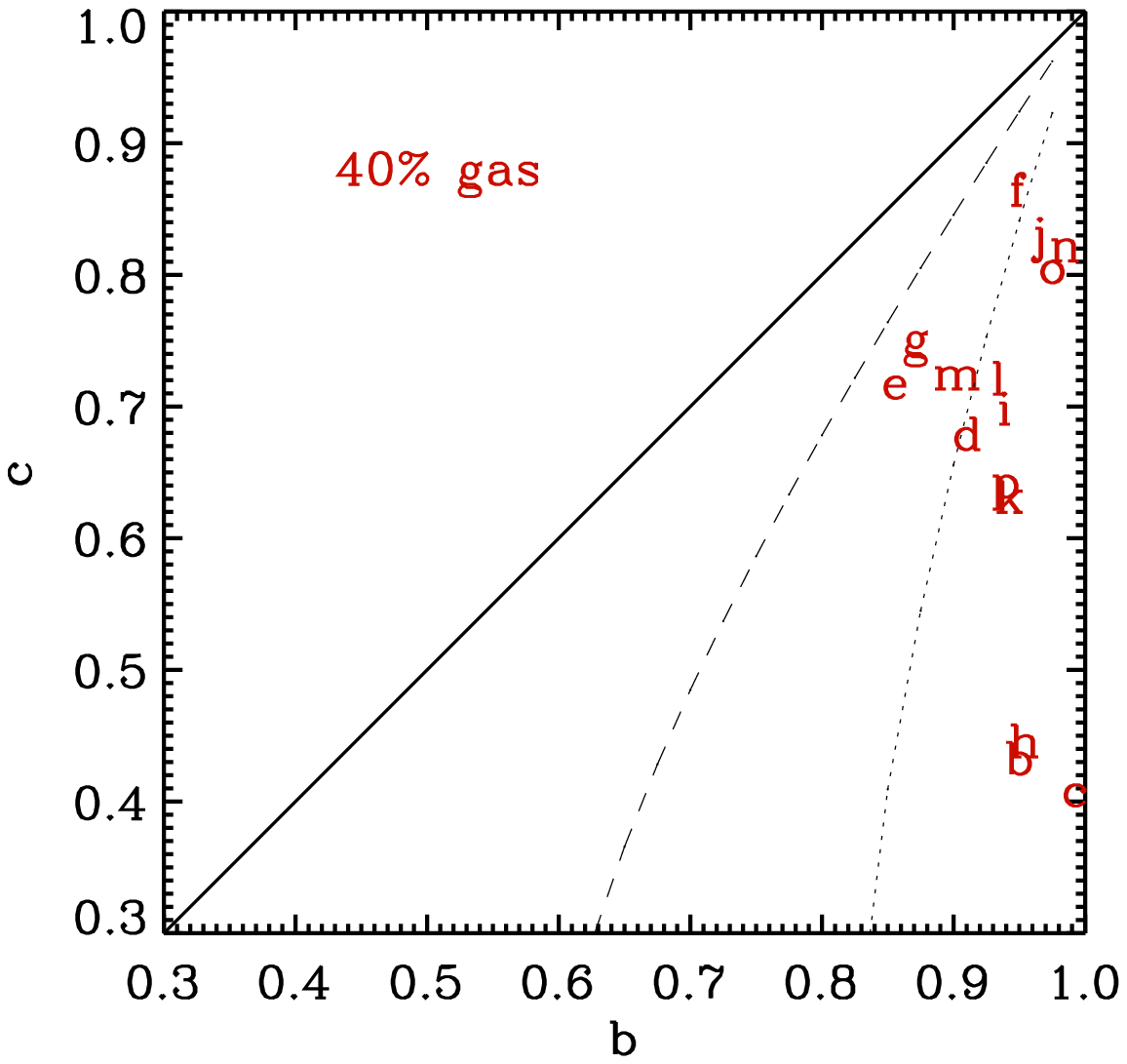}{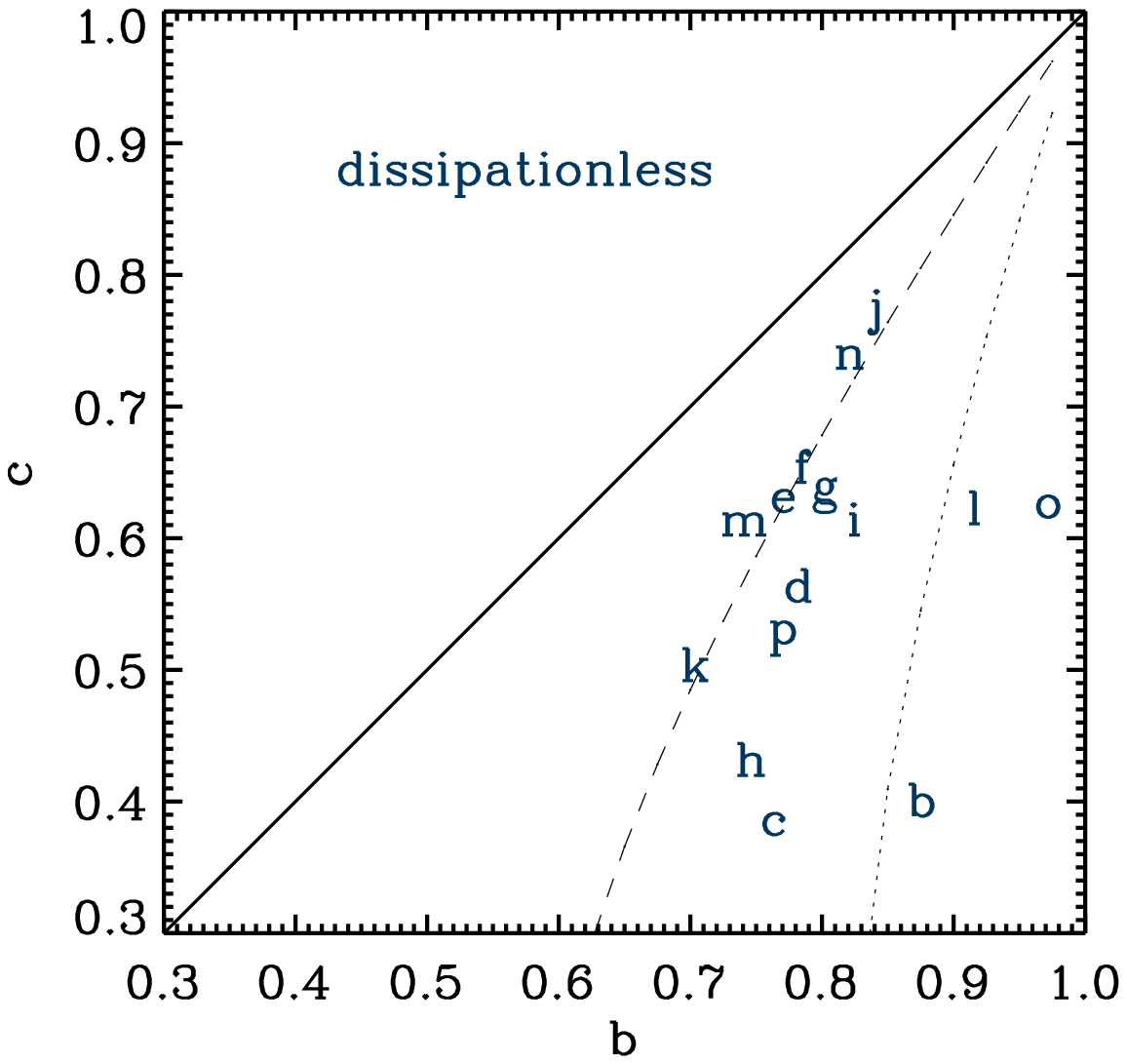}
\caption{The shape diagram for merger remnants.  
$b$ and $c$ are defined as
$(\lambda_2/\lambda_1)^{1/2}$, and $(\lambda_3/\lambda_1)^{1/2}$,
respectively.  $\lambda_1$, $\lambda_2$, and $\lambda_3$ are 
the eigenvalues of the inertia tensor in ascending order.
Only the most bound half of particles are used when computing
the inertia tensor.  The solid line which denotes $c=b$ would
be a prolate spheroid with $T=1$, where $T=(1-b^2)/(1-c^2)$ is 
the triaxiality parameter.  The right vertical axis, at $b=1$
would be an oblate spheroid with $T=0$.  The dashed line shows 
$T=2/3$, and the dotted line shows $T=1/3$.
\label{fig:shape}}
\end{figure*}

\begin{figure}
\plotone{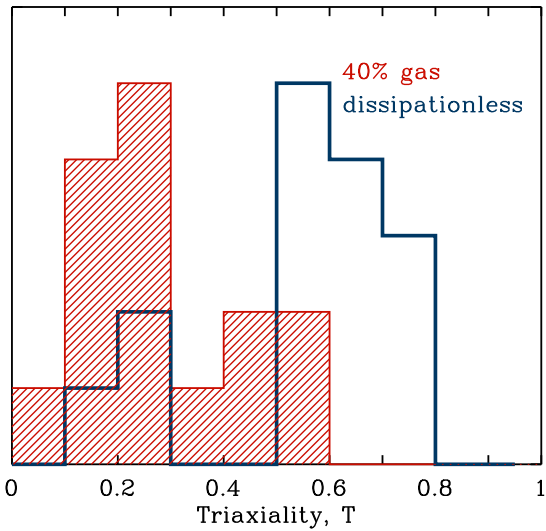}
\caption{Histogram of remnant ``triaxiality'' parameters $T$, where
$T=(1-b^2)/(1-c^2)$, as defined by \citet{FIZ91}.  Oblate galaxies
have $T=0$, prolate galaxies have $T=1$, and all values in between,
as the remnants presented here, are triaxial.
\label{fig:triax}}
\end{figure}

While Table~\ref{tab:shapes} is useful for assessing the effects of
dissipation for any particular merger, the overall trends are more
apparent if one plots the axial ratio $a$ against $b$ as is done in
Figure~\ref{fig:shape}, or a histogram of the triaxiality parameter
$T$ as in Figure~\ref{fig:triax}.  Both of these figures show that
dissipationless remnants are triaxial with a tendency to be more
prolate.  This result is also apparent in previous studies of
dissipationless mergers \citep{B92,HRemI,Sp00}.

Figures ~\ref{fig:shape} and \ref{fig:triax} demonstrate that
dissipational remnants are also triaxial, however these remnants tend
to be much closer to oblate.  This result is consistent with previous
simulations that contained 20\% gas \citep{Sp00}, or a dense stellar
bulge \citep{HRemII,Sp00,GG05a}.  It is noteworthy that the three
dissipational systems that appear to be oblate isotropic rotators
($d$, $f$, and $i$) from their swath on the anisotropy diagram are
actually quite triaxial ($T=0.34, 0.41$, and 0.24, respectively).
However, the shape of merger remnants is known to be a function of
radius \citep{B92,HRemI,HRemII}, and by simply repeating our analysis on
particles which are progressively more bound, we find that the central
regions of the remnants are much more oblate than the outer regions.
Therefore, the apparent discrepancy between the rotation of the
remnant and its shape is likely a byproduct of our selecting the most
bound half of stellar particles rather than a genuine sign of 
anomalous kinematics.  In
\S~\ref{ssec:dep} we investigate how our measure of the
triaxiality depends on orbit, gas fraction and progenitor mass.

Because real ellipticals can be observed from only a single projection,
their intrinsic shapes are difficult to assess.  From the distribution of
observed axial ratios, several works have shown that elliptical galaxies
can not be entirely composed of oblate \citep{Ryd92,LML92,TM95,AR02} or
prolate \citep{TM95} spheroids.  \citet{MT96} presented intriquing evidence
that there exists a bimodal distribution of shapes, namely low luminosity 
ellipticals are consistent with being oblate spheroids, while bright 
elliticals are only consistent with a triaxial intrinsic shape that is
rounder on average than low-luminosity elliticals.  A similar conclusion
was reached by \citet{FIZ91} from analysis of a much smaller set of 
ellipticals that also contained kinematic information.  In their study,
the triaxial shapes are necessary to match the small percentage of galaxies
with rotation observed along their photometric minor axis (see
\S~\ref{ssec:minrot} for additional details) and we note that their 
mean triaxiality parameter $T$ was less than 0.4.  We find that the
mean triaxiality parameter for the dissipational remnants is 0.28 as
opposed to 0.55 for the dissipationless remnants.

%

\begin{figure*}
\plottwo{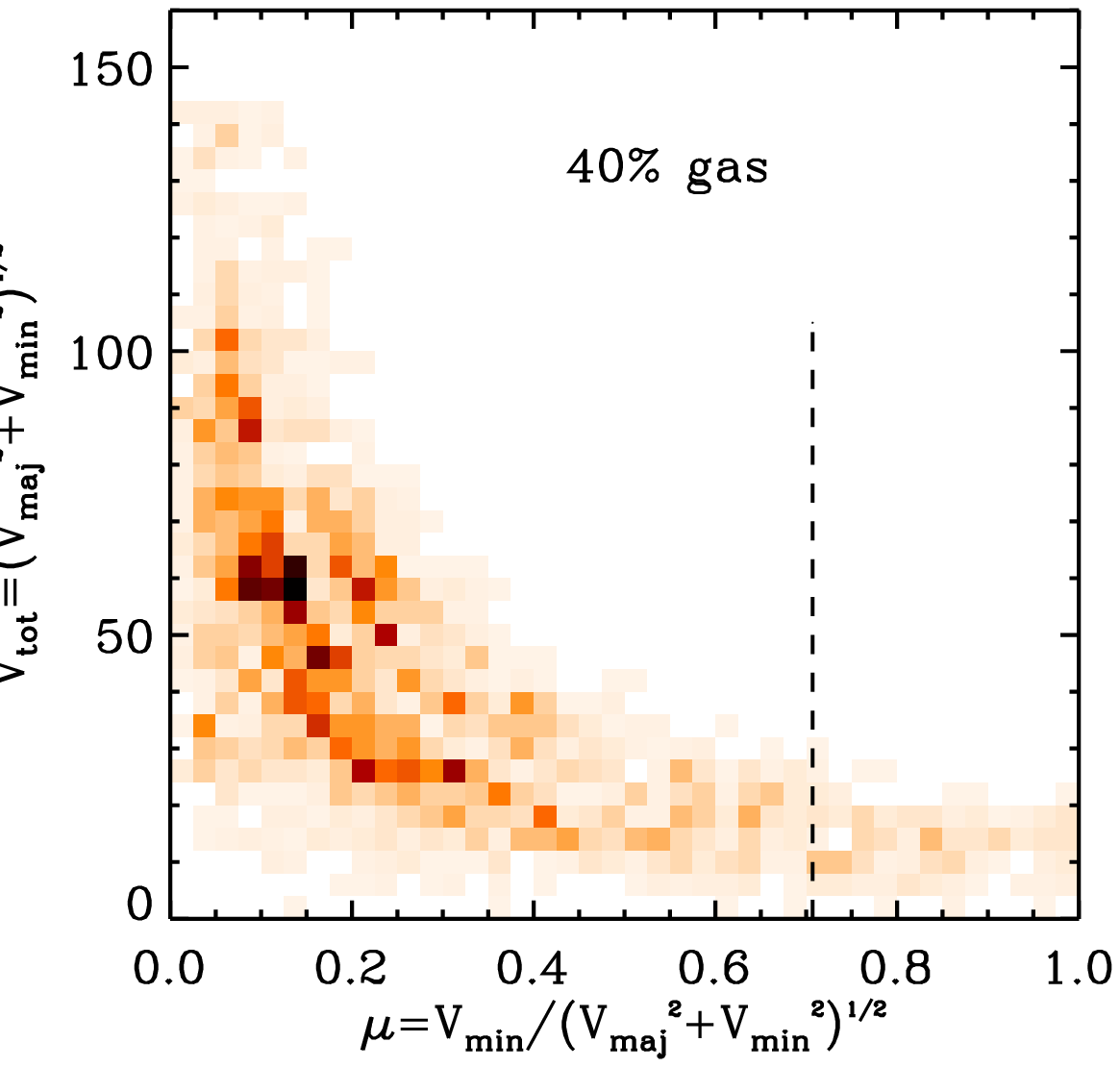}{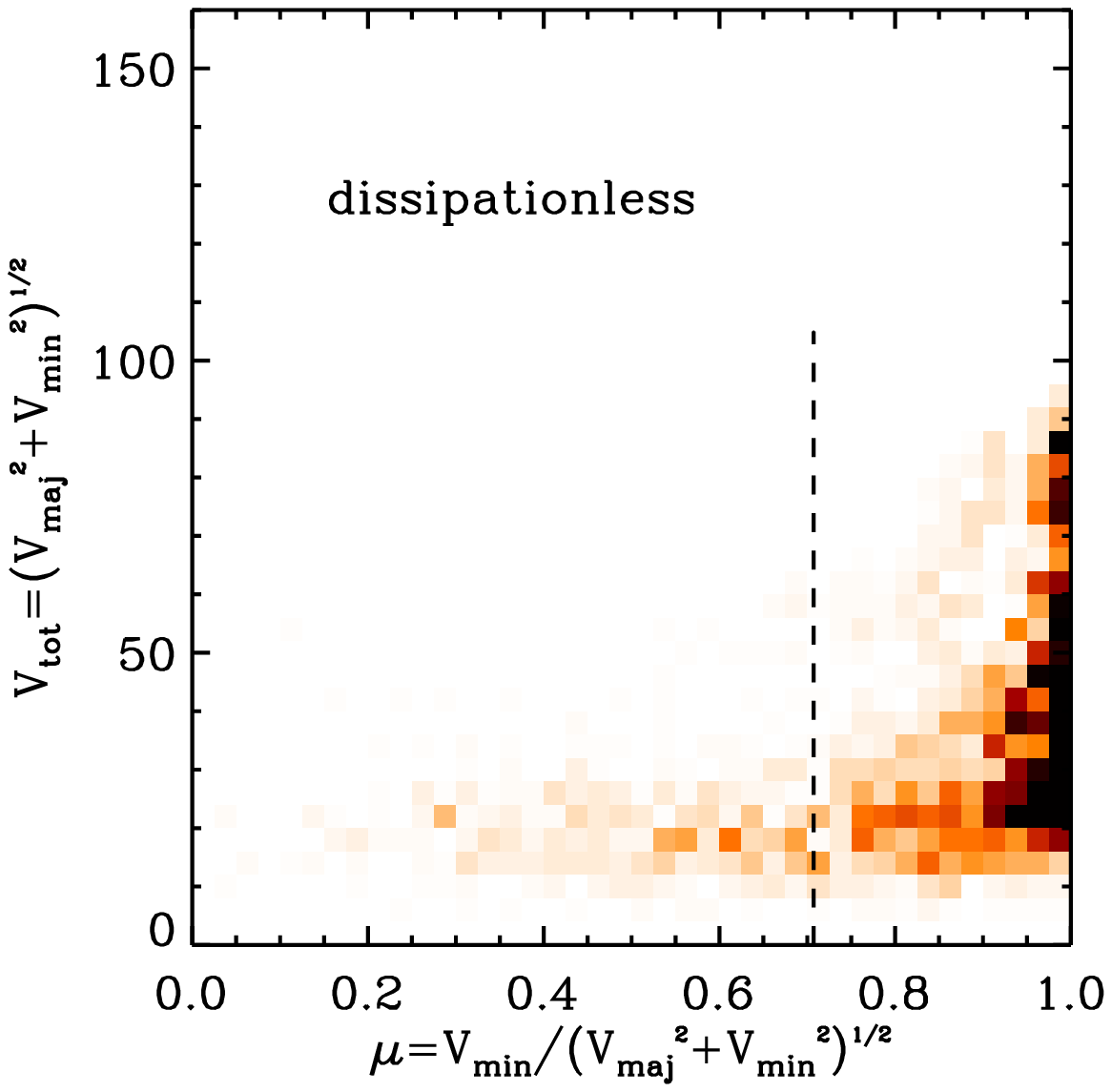}
\caption{The total rotation velocity plotted against the minor axis
rotation parameter, $\mu$, as defined by Equation~\ref{eq:mu}.  The
dashed vertical line denotes equivalent rotation along the major
$V_{\rm maj}$ and minor $V_{\rm min}$ axes.
\label{fig:vsig_min}}
\end{figure*}

\subsection{Minor Axis Rotation}
\label{ssec:minrot}

A number of elliptical galaxies are observed to have rotation along
their minor axes \citep[see e.g.,][and references therein]{FIZ91}.
The presence of minor axis rotation is possible in a triaxial or
prolate ellipsoid and impossible in an oblate, axisymmetric one
\citep{Bin85,dZF91}.  The previous section demonstrated that all of
our merger remnants show some degree of triaxiality and hence it is
possible that any, or all, or our remnants could have some degree of
minor axis rotation.  However, because the dissipationless remnants
are much more prolate (and definitely more triaxial) than 
dissipational ones, it is likely that
dissipationless remnants will have more minor axis rotation.

In order to determine the degree of minor axis rotation, we repeat the
analysis outlined in \S~\ref{ssec:anal} with the slit now placed
along the minor axis of each remnant.  Thus, for each remnant we have a
rotation speed along the major ($V_{\rm maj}$) and minor ($V_{\rm
min}$) axes.  We list the minor axis rotation for each of our merger
remnants in Table~\ref{tab:minr}.  As an additional measure of the
minor axis rotation we follow \citet{Bin85} and measure
\begin{equation}
\mu = 
\frac{V_{\rm min}}{(V_{\rm maj}^2 + V_{\rm min}^2)^{1/2}},
\label{eq:mu}
\end{equation}
the minor axis rotation parameter.  An additional measure of the minor
axis rotation, and one that is useful because it can be compared
directly to observations, is the kinematic misalignment $\Psi$, defined
by
\begin{equation}
{\rm tan}\Psi = \frac{V_{\rm min}}{V_{\rm maj}}.
\label{eq:psi}
\end{equation}
Both minor axis parameters, $\mu$ and $\Psi$, as well as the minor
axis rotation speed $V_{\rm min}$, are listed in Table~\ref{tab:minr}
for each merger remnant.

Surprisingly, Table~\ref{tab:minr} shows that the dissipationless
and dissipational remnants have nearly {\it identical} minor axis
rotation speeds.  However, because the {\it major} axis rotation is
much larger for the dissipational remnants, both $\mu$ and $\Psi$ are
much smaller for the dissipational merger remnants.

\begin{deluxetable}{c|rcc|rcc}
\tabletypesize{\scriptsize}
\tablecaption{Minor Axis Kinematics\label{tab:minr}}
\tablewidth{0pt}
\tablehead{
\colhead{} &
\multicolumn{3}{c}{Dissipationless} &
\multicolumn{3}{c}{40\% Gas}\\
\colhead{ID} &
\colhead{$V_{\rm min}$} &
\colhead{$\mu$} &
\colhead{$\Psi$} &
\colhead{V$_{\rm min}$} &
\colhead{$\mu$} &
\colhead{$\Psi$}\\
\colhead{} & 
\colhead{(\kms)} & \colhead{} & \colhead{} & 
\colhead{(\kms)} & \colhead{} & \colhead{}
}
\startdata
h & 11.2 & 0.59 & 39.3 & 11.3 & 0.30 & 18.1 \\
b & 10.6 & 0.64 & 41.9 &  8.2 & 0.37 & 23.0 \\
c & 10.0 & 0.52 & 33.1 &  8.3 & 0.10 & 6.2 \\
d & 30.1 & 0.93 & 71.1 & 29.2 & 0.29 & 18.0 \\
e & 14.9 & 0.78 & 55.0 & 12.6 & 0.70 & 47.3 \\
f & 66.9 & 0.93 & 74.5 & 32.8 & 0.23 & 13.7 \\
g & 14.5 & 0.70 & 47.3 & 14.4 & 0.31 & 20.7 \\
i & 43.6 & 0.90 & 68.3 & 31.1 & 0.31 & 19.3 \\
j & 22.9 & 0.77 & 53.8 & 34.9 & 0.50 & 32.3 \\
k & 19.1 & 0.87 & 63.5 & 19.5 & 0.32 & 19.8 \\
l & 44.6 & 0.82 & 57.9 & 52.7 & 0.51 & 32.3 \\
m & 52.6 & 0.94 & 73.8 & 54.9 & 0.57 & 37.0 \\
n & 35.6 & 0.91 & 68.8 & 37.1 & 0.33 & 20.4 \\
o & 40.0 & 0.75 & 51.1 & 40.3 & 0.36 & 23.0 \\
p & 32.1 & 0.94 & 74.0 & 22.6 & 0.34 & 20.6 \\
\enddata
\tablecomments{Minor axis rotation properties of
the simulated merger remnants. $V_{\rm min}$ is
the minor axis rotation speed, in \kms, $\mu$ 
is the minor axis rotation parameter, defined by
Equation~\ref{eq:mu}, and $\Psi$ is the kinematic
misalignment defined by Equation~\ref{eq:psi}.
}
\end{deluxetable}

Figure~\ref{fig:vsig_min} presents another way to visualize the
relation between minor and major axis rotation by plotting the minor
axis rotation parameter $\mu$ against the ``total'' rotation speed
$V_{\rm tot}=(V_{\rm maj}^2 + V_{\rm min}^2)^{1/2}$.  All merger
remnants inhabit a clear band in the $\mu$-$V_{\rm tot}$ plane.
However, the dissipational and dissipationless remnants are
concentrated in very different regions.  Fast rotators, those with
$V_{\rm tot}> 50$~\kms, have very little minor axis rotation and are
produced solely by dissipational mergers.  Slow rotators, those with
total rotation speeds below 50~\kms have a wide variety of minor axis
rotation parameters.  Also, there is a clear preference for
dissipationless merger remnants to have significant minor axis
rotation relative to the total rotation.  In fact, the majority of the
dissipationless merger remnants have more minor axis rotation than
major axis rotation.

\begin{figure}
\plotone{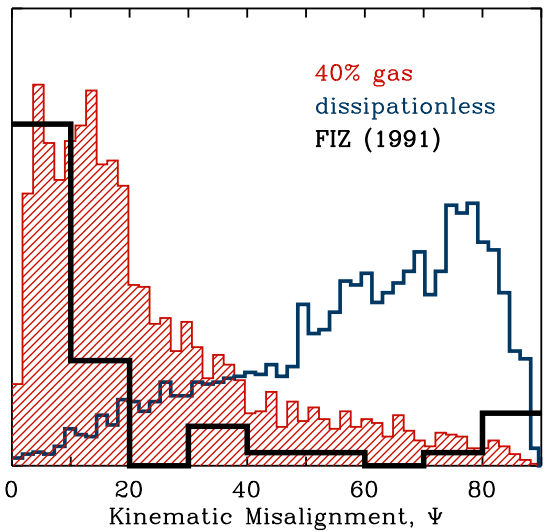}
\caption{Histogram of the kinematic misalignment angle
$\Psi$.  As with the previous figures, the filled 
histogram is for dissipational
remnants and the open histogram is for dissipationless remnants.
Overplotted is data from 44 ellipticals compiled by
\citet{FIZ91}.
\label{fig:kma}}
\end{figure}

To confirm the large difference between the minor axis kinematics of
dissipationless versus dissipational merger remnants, we show the
histograms of the kinematic misalignment parameter $\Psi$ from all
projections in Figure~\ref{fig:kma}.  Again, a clear dichotomy in
$\Psi$ is shown.  Dissipationless remnants have significant minor axis
rotation and thus large kinematic misalignments.  Overplotted are data
compiled by \citet{FIZ91} shown with a thick black histogram.  While
it is clear that there are elliptical galaxies with observed minor
axis rotation, the majority (77\%) have $\Psi<20^\circ$.  There seems
to be a large peak of galaxies with little or no observed minor axis
rotation, $\Psi\sim0$, and another, much smaller, peak at
$\Psi\sim90$, although the latter contains only four galaxies, and
thus its significance is difficult to assess.  In any case, the
observed distribution of kinematic misalignments is much better
represented by the remnants produced in dissipational mergers.

%
\subsection{Isophotal Shapes}
\label{ssec:a4}

The isophotes of elliptical galaxies are not perfect ellipses.  The
distortions from a perfect ellipse can be quantified by expanding 
the residuals from the fitted ellipse in a Fourier series,
\begin{equation}
\Delta r = \sum_k \left[ a_k cos(k\phi)
    + b_k sin(k\phi)\right],
\end{equation}
where $\Delta r$ are the deviations from a perfect ellipse as 
a function of angle $\phi$.  The coefficient $a_4$ of the $cos(4\phi)$
term measures deviations symmetric to the major axis.  Positive
values of $a_4$ indicate disky isophotes and negative values 
indicate boxy isophotes \citep{Ben88,BDM88}.

As mentioned in the introduction, elliptical galaxies appear
to come in two types, either disky, or boxy.  This dichotomy
led \citet{KB96} to suggest a modified Hubble classification
system in which ellipticals are delineated by their isophotal
shapes as opposed to the standard ellipticity.  One line of 
speculation suggests that disky isophotes are an indication of
increased dissipation, and thus the dichotomy in isophotal
shapes represents varying amounts of dissipation during a merger
\citep{BBF92}.  However, numerical simulations have shown that 
remnants formed from dissipationless merging can appear both boxy 
and disky, depending on the viewing angle and isophotal radius
\citep{HRemIV}.  It has also been shown that there is a weak 
trend for dissipational remnants to be more disky 
\citep{Sp00,BS97} probably owing to the production of central 
density cusps that destabilize box orbits \citep{GB85,MF96,BH96}.

\begin{deluxetable}{c|c|c}
\tabletypesize{\scriptsize}
\tablecaption{Isophotal Shapes\label{tab:isophot}}
\tablewidth{0pt}
\tablehead{
\colhead{} &
\colhead{Dissipationless} &
\colhead{40\% Gas}\\
\colhead{ID} &
\colhead{$100 a_4/a$} &
\colhead{$100 a_4/a$}
}
\startdata
h & -1.5 &  0.9  \\
b &  0.3 &  1.2  \\
c & -1.8 &  1.6  \\
d & -1.3 &  1.8  \\
e & -1.0 &  0.4  \\
f &  1.0 &  0.9  \\
g &  2.5 &  0.9  \\
i &  0.1 &  1.9  \\
j &  0.5 &  0.1  \\
k & -2.5 &  0.8  \\
l &  2.7 & -0.3  \\
m &  0.4 & -0.8  \\
n &  0.0 &  1.6  \\
o & -0.9 &  1.7  \\
p & -0.2 &  1.0  \\
\enddata
\tablecomments{Average isophotal shape parameter $a_4/a$
multiplied by one hundred
for each simulated merger remnant.  Positive values
of $a_4/a$ denote disky isophotes, while negative
values denote boxy isophotes.
}
\end{deluxetable}

We measure the deviations from perfect ellipses for each of 
our remnant images and list the average for each orientation
in Table~\ref{tab:isophot}.  From these averages we confirm
that the dissipational remnants are more disky, in general,
than their dissipationless counterparts.  We do caution,
however, that the scatter within any particular remnant is 
large, reflecting that remnants can appear disky or boxy 
depending on the viewing angle.

\begin{figure}
\plotone{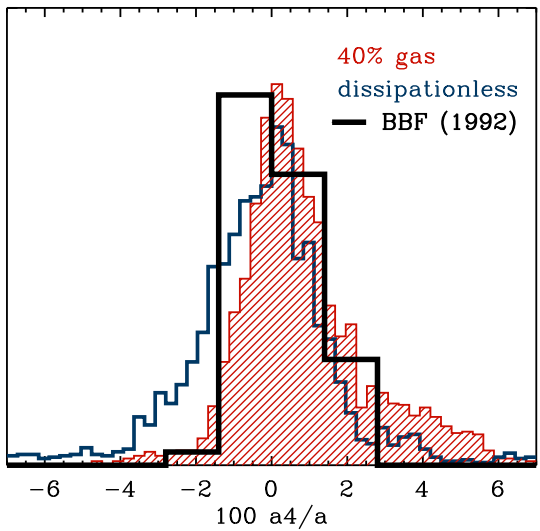}
\caption{Histogram of 100$a_4/a$, which measures the isophotal
deviations from a perfect ellipse; positive values indicate
disky isophotes and negative values indicate boxy isophotes.
As with the previous figures the filled histogram is for
dissipational merger remnants and the open histogram is for
the dissipationless merger remnants.
Overplotted is data from 59 ellipticals compiled by
\citet{BBF92}.
\label{fig:a4}}
\end{figure}

In Figure~\ref{fig:a4} we show a histogram of the coefficient
$a_4$ (normalized by the semi-major axis $a$ and multiplied by
an arbitrary factor of one hundred).  Also included in this figure is data
from 59 ellipticals compiled by \citet{BBF92}.  In general, there
is good agreement between the simulated merger remnants and the
data.  While most isophotes having relatively small deviations from
perfect ellipses, dissipational remnant tend to be slightly disky
and dissipationless remnants slightly boxy.  Observationally, there
are an equal number of disky and boxy remnants.

We note that there exists a tail of disky dissipational remnants and
boxy dissipationless remnants not present in the data.  This discrepancy
may be a result of resolution, as degrading the resolution of our
image, or smoothing it, both reduce the measured isophotal deviation.
We also note that many of the disky dissipational remnants are the fast
rotating systems viewed edge-on.

\begin{figure*}
\plottwo{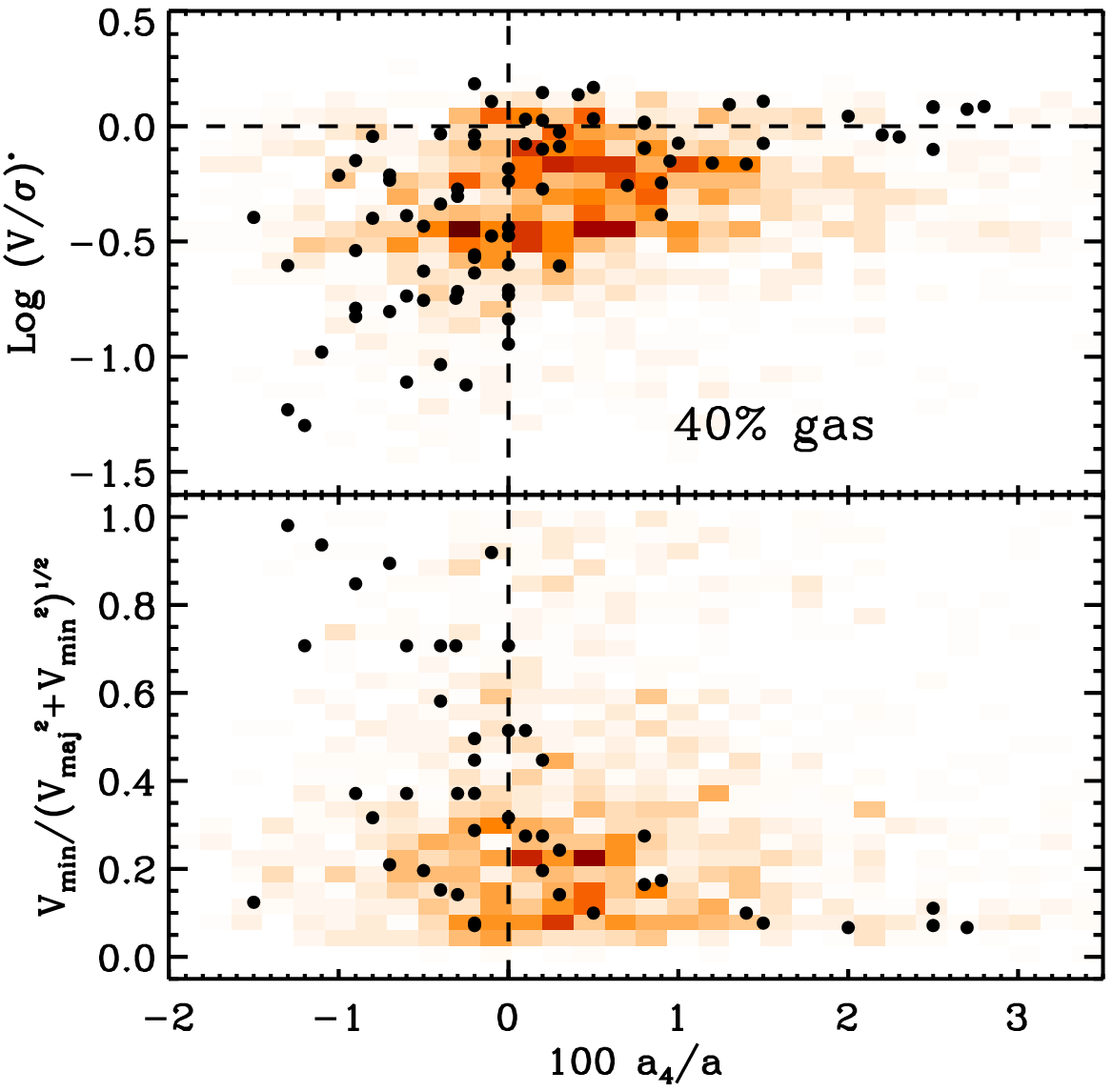}{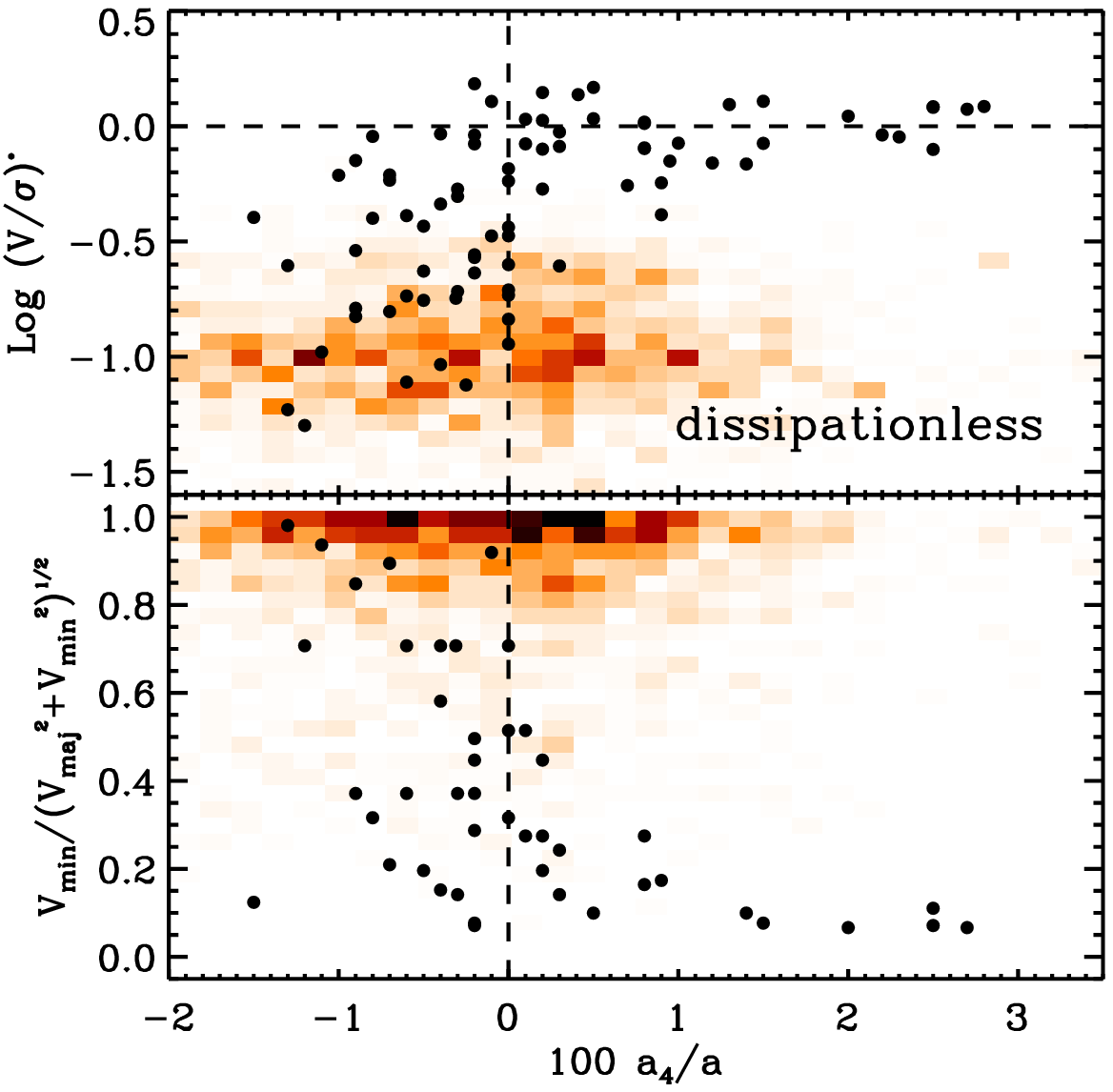}
\caption{A density map that displays correlations between
isophotal shape and rotation.  100$a_4/a$, which measures the isophotal
deviations from a perfect ellipse; positive values indicate
disky isophotes and negative values indicate boxy isophotes.  The
top panel shows the rotation parameter ($V/\sigma$)$^*$ and
the lower panel shows the minor axis rotation parameter of 
Eq.~\ref{eq:mu}.
Overplotted is the identical data as presented in \citet{KB96}.
\label{fig:a4maps}}
\end{figure*}

As evidence for the dichotomy between disky and boxy ellipticals,
\citet{KB96} presented correlations between the $a_4$ parameter
and ($V/\sigma$)$^*$ and minor axis rotation parameter $\mu$.
These correlations are presented in Figure~\ref{fig:a4maps}.
Neither series of remnants presents a clear dichotomy between
disky/boxy and rotation, as is seen in the data.  The 
dissipationless remnants, in particular, appear to be an 
especially poor match to the data as there are a significant number
of slow, ($V/\sigma$)$^* \ll 1$, and minor axis, $\mu  \sim 1$,
rotators with disky isophotes.  Elliptical galaxies with these
rotational properties are {\b all} boxy, in disagreement with the
simulations.

Overall, the dissipational remnants are a much better match to
the general trends present between the rotation and isophotal
shape of elliptical galaxies.  Most of these remnants have very
little minor axis rotation, $\mu < 1$, ($V/\sigma$)$^* \approx 1$,
and have isophotes that are both disky and boxy.  There are a
large number of observational analogs to these systems.  However,
Figure~\ref{fig:a4maps} makes it evident that we are still unable
to produce slow and minor axis rotators that are uniformly boxy.
This will be discussed further in \S~\ref{sec:disc}.

%
\subsection{Dependencies}
\label{ssec:dep}

The previous sections demonstrate that there is a significant
difference between the size, shape and rotational properties of merger
remnants when the initial disk contains 40\% gas compared to when the
entire disk is dissipationless.  However, it is unclear whether these
results are sensitive to the various assumptions implicit to our
merger simulations.  While a complete exploration of the large
parameter space that describes the disk galaxy models and their
encounters is beyond the scope of this work, we can provide some
indications of the robustness of our results by systematically varying
several parameters which we suspect may affect the properties of the
merger remnants.  Specifically, the following three sections consider
variations in the progenitor disk gas fraction, the orbital angular
momentum, and the size (mass) of the progenitor disk galaxies.

For simplicity we select the $e$, $h$, and $k$ disk orientations
from Table~\ref{tab:orbs} to resimulate throughout this section.  
These three are selected to span a range of remnant properties.

\subsubsection{Dependence on Gas Fraction}
\label{sssec:dep-gf}

\begin{figure}
\plotone{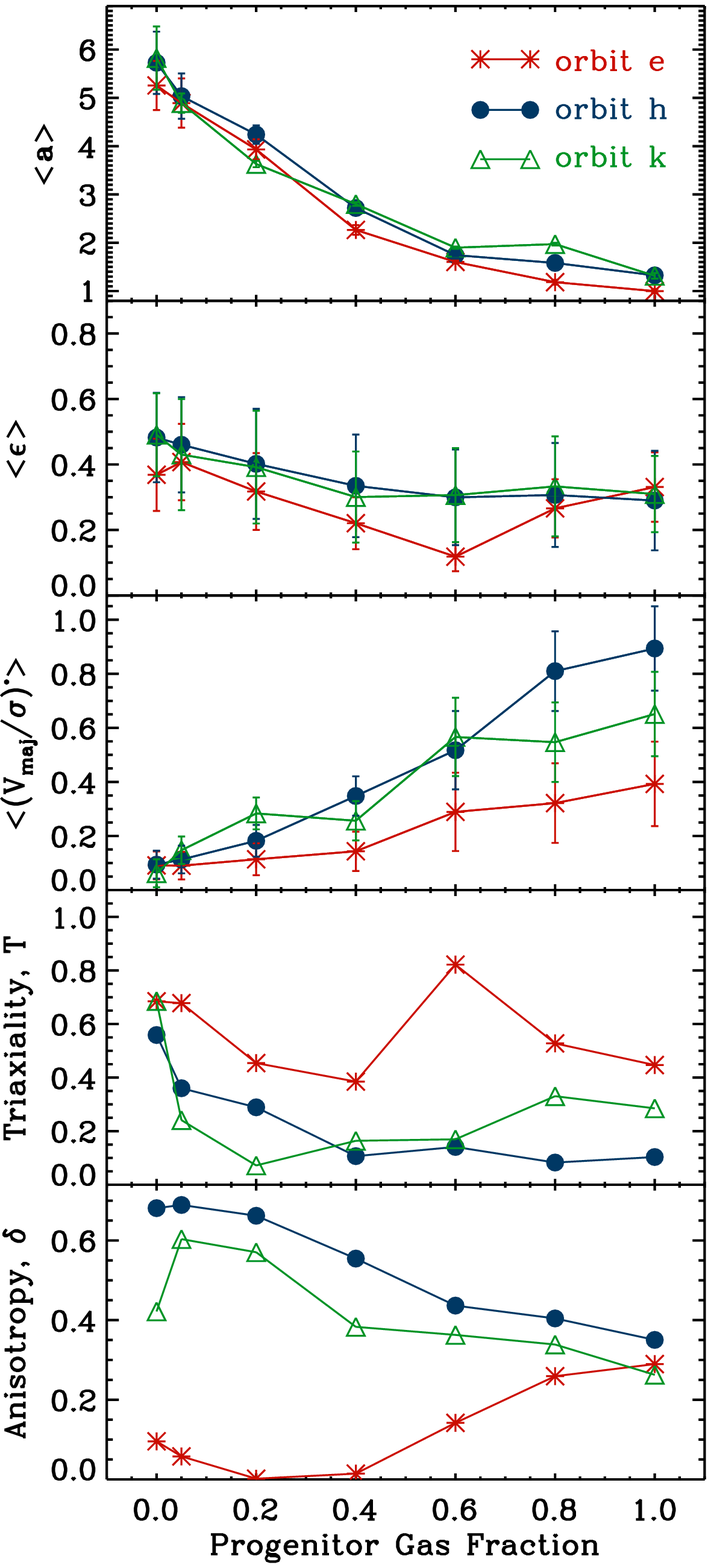}
\caption{The semi-major axis $a$, ellipticity $\epsilon$, 
and V$_{\rm maj}$/$\sigma$ averaged over 190 different 
lines of sight and the triaxiality parameter $T$, and
anisotropy parameter $\delta$ for our $e$, $h$ (prograde-prograde),
and $k$ orientation merger remnants as a function of the gas 
fraction of the progenitor disk.  A key is provided in
the top panel which denotes the symbols used for the 
three initial disk orientations.  For the top three
panels, which present quantities averages over many lines
of sight, one sigma error bars are included to provide a
sense of the variation introduced by projection effects.
\label{fig:gf_dep}}
\end{figure}

To begin, we systematically change the disk gas fraction from zero,
i.e., the dissipationless case, to a pure gas disk.  For reference,
our fiducial dissipational simulation contained a disk with 40\% gas.
In Figure~\ref{fig:gf_dep} we show the size, as measured by the
semi-major axis, ellipticity, and ($V/\sigma$)$^*$, all averaged over
all projections for each merger remnant.  In addition, we show the
triaxiality parameter $T$, and the anisotropy $\delta$ for each
remnant.

As might be expected from the results of \S~\ref{sec:results},
there is a systematic trend for remnants produced by the collision of
spiral galaxies with an increasing fraction of gas to be smaller,
rounder, and to have more rotation.  However, some of the remnant
properties are more sensitive to the gas fraction than others.  For
example, the size of remnants shrinks and ($V/\sigma$)$^*$ steadily
rises with increasing gas fraction.  But, the projected ellipticity
becomes only slightly rounder, and at a level that is comparable to
the variations expected from projection effects alone (shown by the
error bars on each point). For each individual orientation the shape,
as measured by the triaxiality parameter $T$, appears to level off
above a gas fraction of 20\%, even though the individual orientations
level off to different shapes.

The dependencies displayed by trends in ($V/\sigma$)$^*$, $T$, and
$\delta$ demonstrate that inherent differences exist purely because of
the initial disk orientations.  For example, the $e$ merger tends to
be much slower rotating, much more triaxial, and contain much less
anisotropy, no matter what the gas fraction is, even though the trends
with gas fraction are similar to those found in $h$ and $k$ mergers.

From this series of 21 simulations (3 orientations, each simulated
with 7 different gas fractions) we conclude that the size and rotation
are strongly affected by the initial disk gas fraction, while the
shape is only moderately affected.  The anisotropy appears more
strongly correlated with the orbital configuration than with the gas
fraction.

These results also directly relate to those of \citet{Rob06fp} who find
that remnants produced in major mergers can reproduce the observed
fundamental plane of ellipticals when the progenitor disks contain gas fractions
greater than 30\%.  Thus, it seems that gas-rich major mergers can
help explain both the kinematics of elliptical galaxies and the tilt
of the fundamental plane.

\subsubsection{Dependence on Orbital Angular Momentum}
\label{sssec:dep-orbit}

\begin{figure}
\plotone{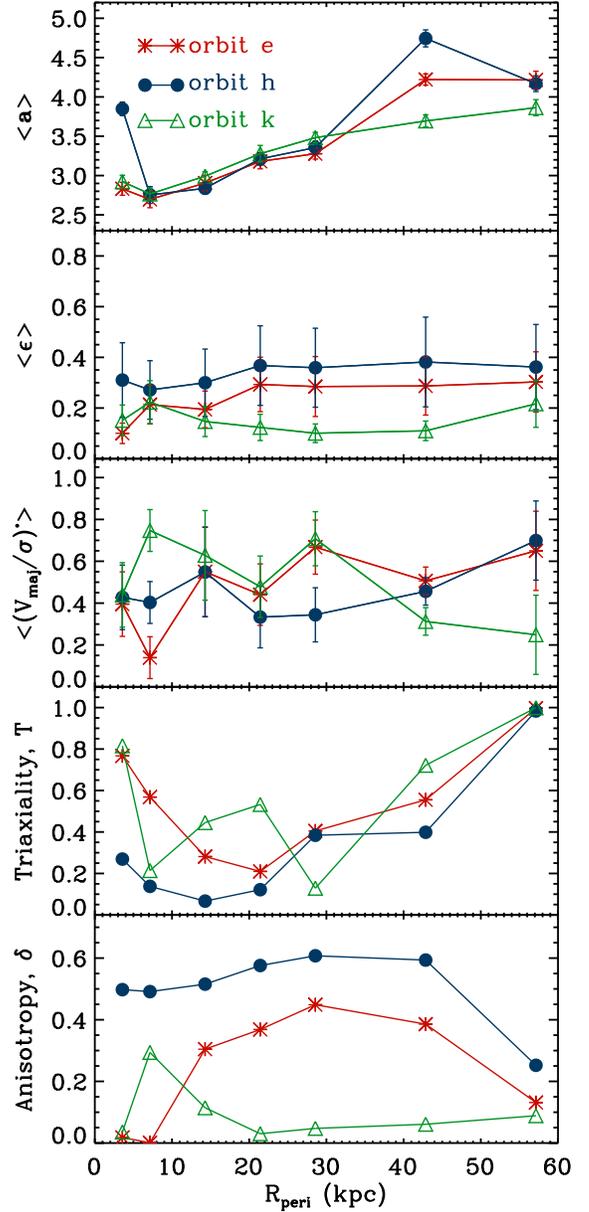} 
\caption{Same as Figure~\ref{fig:gf_dep} except here
the remnant properties are plotted against the
orbital pericentric distance.
\label{fig:orb_dep}}
\end{figure} 

Next, we note that merger orbits as determined by cosmological
simulations span a range of impact parameters \citep{KB01}.  In the
fiducial series of mergers, analyzed in \S~\ref{sec:results}, we
assumed each merger orbit was parabolic, and had a pericentric
distance of 7.1 kpc, generating an orbit with a low impact
parameter.  In this series of runs, we resimulate the dissipational
mergers ($e$, $h$, and $k$) with a range of pericenter separations
between 3.6 and 57 kpc.  The initial starting separation for all
mergers was 143 kpc.  We note that all of these orbits contain an
identical amount of energy (zero) yet have progressively more orbital
angular momentum ($L^2\sim R_{\rm peri}$).

Figure~\ref{fig:orb_dep} is equivalent to Figure~\ref{fig:gf_dep}
except in the present figure the horizontal axis measures the
pericentric distance.  The size of each merger remnant is a moderate
function of $R_{\rm peri}$, with wide orbits producing larger
remnants.  The apparent ellipticity has no discernible correlation
with $R_{\rm peri}$.  The rotation shows weak, or no, correlation to
$R_{\rm peri}$.  At first glance, this lack of correlation between
rotation and orbital angular momentum seems odd.  However, two
effects act to offset the transfer of orbital angular momentum to
remnant rotation.  First, the initial disk galaxies contain massive
dark halos which soak up the majority of the orbital 
angular momentum.  Second, the orbits with high orbital angular
momentum require a longer time to merge, and thus the initial
disks have consumed a larger fraction of their initial gas.  In
this sense, they are effectively lower gas fraction mergers, which
the previous section showed are slower rotators.

As we saw in the last section, the shapes of merger remnants depend on
the original disk orientations.  Figure~\ref{fig:orb_dep} demonstrates
that the shape also depends on $R_{\rm peri}$.  Wide orbits tend to
produce prolate ($T\approx1$) remnants, regardless of the initial disk
orientation, while intermediate and radial orbits produce remnants
that are closer to oblate.  However, the orbits with $R_{\rm
peri}<30$~kpc begin to segregate based upon the initial disk
orientation.  There is a trend, though, for very radial orbits to be
more prolate than intermediate orbits.

There is a mild tendency for very wide orbits to produce more
isotropic ($\delta\sim0$) remnants.  This correlation, however, is
weak, at best, especially compared to the strong dependence on 
the initial disk orientations.

\subsubsection{Dependence on Progenitor Mass}
\label{sssec:dep-mass}

\begin{figure}
\plotone{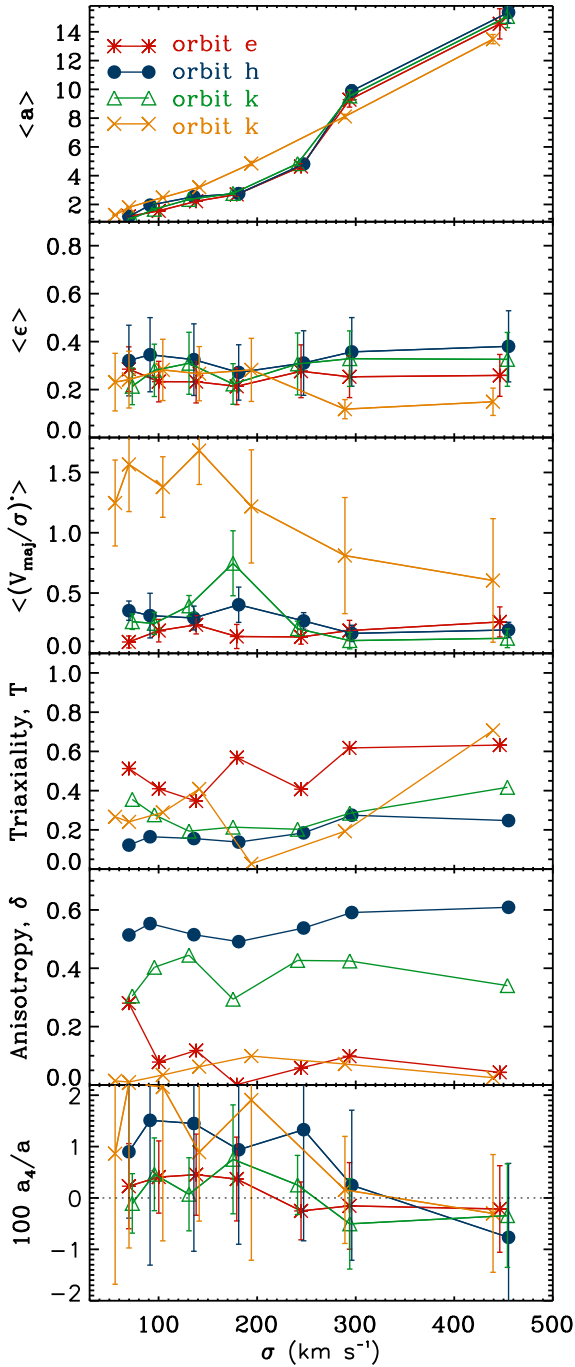}
\caption{Same as Figure~\ref{fig:gf_dep} except here
the remnant properties are plotted against the
central velocity dispersion $\sigma$ of each merger remnant.
\label{fig:mass_dep}}
\end{figure}

As a final dependency we investigate the relation between progenitor
mass and the remnant properties.  In order to
span a wide range of remnant central velocity dispersions, we have
simulated our standard $e$, $h$, and $k$ mergers with initial disk
galaxies both smaller ($V_{200}= 56, 80, 115$~\kms) and larger
($V_{200}= 220, 320, 500$~\kms) than our fiducial case
($V_{200}=160$~\kms).  For these mass excursions, all other progenitor
disk properties are kept fixed, such as gas fraction (40\%), spin
(0.033), dark matter concentration (9), and specific angular momentum
(0.05).  The initial orbits are scaled such that the ratio $R_{\rm
peri}/R_d$ remains constant.  For reference, our fiducial model
($V_{200}=160$~\kms) has a total virial mass of
$1.4\times10^{12}$~\msun, and our mass excursion samples galaxies a
factor of 20 smaller and a factor of 30 larger in mass.

Figure~\ref{fig:mass_dep} shows the resulting remnant properties as a
function of remnant central velocity dispersion.  In general, it
appears as if the size is the strongest dependency, as expected
since $R_d \propto V_{200}$.  We find a
size--mass relationship consistent with \citet{Rob06b,Rob06fp}, who
simulate binary mergers for an equivalent range of progenitor disk
masses.  The ellipticity and rotation do not appear to correlate
with mass.  Once again, the triaxiality and
anisotropy depend on the initial disk orientations more than
on the size of the merging systems.

From our investigation of the remnant properties as a function of
progenitor mass we conclude that mass is of little importance in
fixing the rotational properties of a merger remnant.  Instead, it is
the progenitor gas fraction and merger orbit and orientation which are
the primary determinants in what is left behind from the collision of
two disk galaxies.

\subsubsection{Other Dependencies}

In the three previous subsections we have tried to determine the
degree to which our results are dictated by our initial assumptions.
We conclude that the initial gas fraction is of critical importance to
the structure of a merger remnant.  The orbit and orientation, as well
as the progenitor disk mass are also factors which determine the
remnant properties.  However, there are still a large number of
parameters (bulge size, radial gas profile, halo $\lambda$ and $c$)
which may in principle influence the remnants that have not been 
addressed so far.

We have performed a single simulation where the initial disk contained
a bulge one-third the mass of the progenitor disk.  The fiducial disk
galaxy model did not contain a bulge.  In this case the remnant showed
systematic, but very small differences from the no-bulge fiducial
remnant.  Specifically, because of the dense central bulge, the
remnant is rounder and has a higher central velocity dispersion.  The
no-bulge remnant also has more rotation, but because $\sigma$
increases, the net effect is to lower $V/\sigma$, and drive the
remnant down and to the left in the anisotropy diagram.  The presence
of a bulge increases the rotation because star formation is suppressed
during the early stages of the merger \citep{MH94majm,MH96} and hence the
effective gas fraction at the final merger is larger than in the
no-bulge case.  We expect that any process, or assumption, that
results in a high gas fraction during the final merger will increase
the rotation of the remnant.  Possible mechanisms are the inclusion of
a bulge, less efficient quiescent star formation, or an extended gas
distribution.

We finally note that the presence of a black hole has very little effect to
the results presented here.  The black hole feedback induced
``blowout'' phase is necessary to terminate star formation and thereby
produce red remnants \citep{SdMH05red}, but this process occurs after
the merger has taken place and subsequent to the production of most of
the remnant stars.  In fact, we ran one case that did not 
include black hole feedback, and thus did not produce a galactic wind,
however the remnant stellar mass increased by less than 5\%.  While
these additional stars add to the ``young'' component shown in
Figure~\ref{fig:slitvel}, the overall changes to the remnant kinematics
are much less than projection effects or merger orientations.
Thus we conclude that the black hole may be very important for the 
long-term color evolution, but has little importance in determining the
bulk kinematics of the remnant because the processes that fix the dynamics
have finished before the blowout.

\section{Discussion}
\label{sec:disc}

The previous section presented an analysis of two series of fifteen
merger simulations, one series consisted of dissipationless mergers,
while the second series contained disks composed of 40\% gas, but
identical in every other respect.  In addition, we investigate the
dependence of our results to variations in the original gas fraction,
orbital angular momentum, and progenitor size/mass.  These tests show
that the size, shape and rotation of any merger remnant is a strong
function of the initial disk gas fraction and the orientations of the
disks.  In what follows we will attempt to place these results into a
broader context and relate them to our current understanding of the
formation and evolution of spheroidal galaxies.

\subsection{Forming Low-luminosity Elliptical Galaxies and Bulges}
\label{ssec:smalle}

As outlined in the Introduction, low-luminosity ellipticals and bulges
fall into the class of galactic spheroids that have disky isophotes,
are X-ray faint, have power-law surface brightness profiles and show
significant rotation.  The rotation axis is typically aligned with their
photometric minor axis and they are consistent with being oblate
isotropic rotators.  Thus, asking ``What mechanism forms
low-luminosity ellipticals and bulges?'' may be equivalent to the
question ``What produces oblate isotropic rotators?''

Our results suggest that the merger of equal mass gas-rich disk
galaxies may produce remnants that are oblate isotropic rotators.
Qualitatively, over 25\% (4/15) of the 40\%
gas major mergers yielded an oblate isotropic rotator.  However, it
is unclear if major mergers can produce these objects in the correct
number.  As
indicated by Figure~\ref{fig:vsig_hist}, the dissipational mergers
produce slightly fewer fast rotating, ($V/\sigma$)$^* \geq 1$,
ellipticals than observed.  Does this dearth of fast rotating remnants
pose a problem to the ``merger hypothesis''?

The short answer to this question is no.  The scenario outlined
in the current paper, major mergers between gas-rich disk galaxies,
is likely only one of a number of possible mechanisms to
produce oblate isotropic rotators.
Another possible mechanism, proposed by \citet{NB03}, is the 
dissipationless merger of unequal mass disk galaxies.  Both
\citet{NB03} and \citet{BCJ04,BJC05} show that unequal mass mergers,
specifially those with mass ratios 1:4 or higher, produce
oblate remnants.

While we
consider there to be strong evidence that all mergers should include
progenitor galaxies with some amount of gas, we suspect that the
inclusion of gas will only increase the amount of rotation in the
remnants of minor mergers.  Finally, we note that 
Figure~\ref{fig:gf_dep} shows that the 
frequency of oblate isotropic rotators will increase when initial
disks of higher gas fraction are merged.  In short, it does not 
apear that there are any problems in producing oblate isotropic
spheroids.

It is also noteworthy that not all of the gas is consumed by the merger
event.  In general, $\sim10-20$\% of the original gas mass remains in the
remnant, typically in a hot phase that is spread throughout the dark
halo \citep{Cox04,Cox06x}.  The cooling time of this gas is several Gyr, and
thus, depending on the angular momentum of this gaseous material, and
the amount of pristine inter-galactic material that is newly accreted,
it is possible a new gas disk will from around, or within, the
existing spheroidal stellar remnant.  In this sense the remnants
produced here may be the precursor to bulges in present day spirals or
at least a hidden disk component.

\subsection{Forming Large Elliptical Galaxies}
\label{ssec:bige}

Luminous elliptical galaxies have boxy isophotes, contain halos of hot
gas, have surface-brightness profiles with cores and demonstrate little
or no rotation.  While forming low-luminosity ellipticals was equivalent
to forming oblate isotropic rotators, an analogous equation does not exist
for luminous ellipticals.  We do note that the primary conclusion of our
work is that dissipational merger remnants are a much better match
to the entire class of elliptical galaxies.  The evidence for this comes
from the distribution of ellipticities (Figure~\ref{fig:allhist}) and
($V/\sigma$)$^*$ (Figure~\ref{fig:vsig_hist}), and the amount of minor 
axis rotation (Figure~\ref{fig:kma}), as measured by $\Psi$, the
kinematic misalignment as well as the correlations between rotation 
and isophotal shape (Figure~\ref{fig:a4maps}).  However, as the isophotal
analysis demonstrates, it is unclear that any of our remnants, dissipational
or dissipationless, are really slow rotators {\it and} uniformly boxy
as it appears all luminous ellipticals are.  This difficulty with
the model brings into question whether luminous ellipticals can be
produced within the merger hypothesis.

While a single gas-rich 
merger appears to well reproduce the properties of low-luminosity
ellipticals, it is possible that luminous ellipticals require a
more complicated merger history.  
In fact, within the hierarchical build-up of
structure predicted by Lambda cold dark matter, we expect mergers to
occur frequently, especially minor mergers, and there is no reason to
believe that galaxies participate in only one major merger.  Recent
observations suggest that, on average, somewhere between one-half and
all elliptical galaxies undergo a spheroid-spheroid merger
\citep{vD05,Bell05} below redshift one.  In the context of forming
bright cluster galaxies, multiple dry mergers has long been a
well-motivated mechanism \citep{Mer85,Dub98}.  As a brief test of the
effects of continued merging, we performed one re-merger, where two
merger remnants were merged in a manner similar to our fiducial series.
In this one case, the remnant rotation and ellipticity decreased
slightly, while the central velocity dispersion increased slightly
and the isophotes became increasingly boxy.
Although it is difficult to say anything concrete from a single
simulation, we speculate that spheroid-spheroid merging will move
remnants vertically downwards in the anisotropy diagram to the region
occupied by luminous slowly-rotating ellipticals while also making
the system uniformly boxy.  A similar result
has been found for mergers of dissipationless remnants by
\citet{NKB06} and we note that other work has also shown
that multiple merger remnants can remain on the fundamental plane
\citep{CdCC95,NLC03,GG03,BK06,Rob06fp} and may have $R^{1/4}$ 
surface brightness profiles
\citep{BJC05}.  However, it is clear that much more work needs to be
performed in order to understand the relationship between different
merger histories and the kinematics of the remnant.

\subsection{Future Considerations}
\label{ssec:future}

As a final point of discussion we note that a considerable amount of
work still needs to be performed in order to prove that the simulated
merger remnants are bona fide elliptical galaxies.  The present study
has shown that dissipational merger remnants appear to be a good 
match (at least much better than dissipationless remnants) to the
ellipticities, ($V/\sigma$)$^*$, and kinematic misalignments $\Psi$
observed in elliptical galaxies.  However, we have not addressed the
surface brightness profiles, metallicities and abundance ratios,
and kinematic subsystems of our merger remnants.  Work toward this end
is currently underway and is necessary as many of these features
are observed to be correlated with the rotational properties suggesting
that all of these properties are a result of a common formation mechanism.

\section{Conclusions}
\label{sec:conclusions}

In this paper we have analyzed a series of numerical simulations to
investigate the kinematic structure of galaxies formed from the
collision of equal-mass disk galaxies.  In particular, we 
determine that remnants produced by the collision of gas-rich disk
galaxies are smaller, rounder, have higher central velocity
dispersions (on average) and larger rotation speeds than the
remnants of dissipationless disk galaxy mergers.

The larger rotation present in gas-rich merger remnants owes its
origin to dissipation.  Stars formed during the merger rotate faster
than stars present before the merger began.  Dissipation and star
formation also produce remnants that are closer to oblate and are
uniformly more isotropic than dissipationless remnants.  Slightly
more than one-quarter of the dissipational merger remnants are
consistent with being oblate isotropic rotators.

Compared to observed ellipticals, the remnants formed from
dissipational mergers are a significantly better match to the
distribution of ellipticities, ($V/\sigma$)$^*$, and kinematic
misalignment angles $\Psi$ of elliptical galaxies than
dissipationless merger remnants.  In particular, dissipationless 
remnants demonstrate significant minor axis rotation and appear
to be flattened much more than observed ellipticals.

We also calculate the isophotal shapes of the simulated merger
remnants.  Dissipational remnants tend to be disky while 
dissipationless remnants tend to be boxy.  Observed ellipticals
are evenly distributed between disky and boxy.  While both
remnants appear to be a sufficient match to the observed 
distribution, when comparing the correlation between disky/boxy
and rotation, the dissipationless merger produce a significant
number of slowly rotating disky remnants where there are no 
observed analogs.  Both dissipational and dissipationless 
mergers produce remnants that are both disky and boxy, and thus
these mechanisms have difficulty reproducing the luminous, slowly
rotating ellipticals that are observed to be uniformly boxy.

In general, our results suggest that dissipationless disk galaxy
mergers {\it cannot} be the dominant mechanism to form elliptical 
galaxies.  Dissipational mergers, on the other hand, appear to be
a viable mechanism to produce elliptical galaxies, specifically
oblate isotropic rotators (i.e., low-luminosity ellipticals), and 
thus our results lend support to the ``merger hypothesis'' provided 
that the progenitor disk galaxies are gas-rich.  As mentioned in
\S~\ref{sec:intro}, additional evidence for the dissipative origin
of ellipticals comes from their high phase space density compared
to spiral galaxies.  Gas dissipation provides a natural mechanism
to increase the phase space density during the merger and also 
appears to be necessary for reproducing the scaling relation of
elliptical galaxies \citep{Rob06fp}.

Our modeling suggests several avenues for further testing this
hypothesis.  Mergers between gas-rich spirals will imprint subtle
features into the remnants.  The central starburst will modify the
inner profiles of the remnants \citep{MH94dsc}, perhaps accounting
central light excesses seen in merging systems \citep{RJ04,RJ06}.
In principle, this can be tested by comparing predictions for
metallicity and color gradients with observations \citep{MH94popg}.
Dissipational merger may also provide a natural mechanism to
produce kinematic subsystems in elliptical galaxies \citep{HB91,
BB00}.  The shells, ripples, loops and other fine structures seen
around many relaxed ellipticals \citep{Schweizer98} are a natural
consequence of mergers involving disk galaxies \citep{HS92}, but
that do not form in major mergers between hot stellar systems.
Determining the ubiquity of fine structure in red galaxies would
further constrain the importance of disk mergers to the formation of
ellipticals.

Placed within the ``cosmic cycle'' of galaxy formation, we can now
argue that gas-rich major mergers trigger quasars and starbursts,
fuel the growth of supermassive black holes, and produce remnant
galaxies which have the colors, scaling relations, and kinematics
akin to present day ellipticals.

\acknowledgements
We thank Marijn Franx for the motivation that initiated this work
and additional comments that significantly improved this paper.
We also thank Ralf Bender for providing the data in Figure~\ref{fig:a4maps}.
This work was supported in part by NSF grants ACI 96-19019, AST
00-71019, AST 02-06299, and AST 03-07690, and NASA ATP grants
NAG5-12140, NAG5-13292, and NAG5-13381.  The simulations were
performed at the Center for Parallel Astrophysical Computing at
Harvard-Smithsonian Center for Astrophysics.

\bibliographystyle{apj}
\bibliography{ms.bbl}

\begin{thebibliography}{105}
\expandafter\ifx\csname natexlab\endcsname\relax\def\natexlab#1{#1}\fi

\bibitem[{{Alam} \& {Ryden}(2002)}]{AR02}
{Alam}, S.~M.~K. \& {Ryden}, B.~S. 2002, \apj, 570, 610

\bibitem[{{Allgood} {et~al.}(2006){Allgood}, {Flores}, {Primack}, {Kravtsov},
  {Wechsler}, {Faltenbacher}, \& {Bullock}}]{All06}
{Allgood}, B., {Flores}, R.~A., {Primack}, J.~R., {Kravtsov}, A.~V.,
  {Wechsler}, R.~H., {Faltenbacher}, A., \& {Bullock}, J.~S. 2006, \mnras, 367,
  1781

\bibitem[{{Barnes}(1988)}]{B88}
{Barnes}, J.~E. 1988, \apj, 331, 699

\bibitem[{{Barnes}(1992)}]{B92}
---. 1992, \apj, 393, 484

\bibitem[{{Barnes} \& {Hernquist}(1996)}]{BH96}
{Barnes}, J.~E. \& {Hernquist}, L. 1996, \apj, 471, 115

\bibitem[{{Barnes} \& {Hernquist}(1991)}]{BH91}
{Barnes}, J.~E. \& {Hernquist}, L.~E. 1991, \apjl, 370, L65

\bibitem[{{Bekki} \& {Shioya}(1997)}]{BS97}
{Bekki}, K. \& {Shioya}, Y. 1997, \apjl, 478, L17+

\bibitem[{{Bell} {et~al.}(2005){Bell}, {Naab}, {McIntosh}, {Somerville},
  {Caldwell}, {Barden}, {Wolf}, {Rix}, {Beckwith}, {Borch}, {Haeussler},
  {Heymans}, {Jahnke}, {Jogee}, {Meisenheimer}, {Peng}, {Sanchez}, \&
  {Wisotzki}}]{Bell05}
{Bell}, E.~F., {Naab}, T., {McIntosh}, D.~H., {Somerville}, R.~S., {Caldwell},
  J.~A.~R., {Barden}, M., {Wolf}, C., {Rix}, H.-W., {Beckwith}, S.~V.~W.,
  {Borch}, A., {Haeussler}, B., {Heymans}, C., {Jahnke}, K., {Jogee}, S.,
  {Meisenheimer}, K., {Peng}, C.~Y., {Sanchez}, S.~F., \& {Wisotzki}, L. 2005,
  ArXiv Astrophysics e-prints

\bibitem[{{Bender}(1988)}]{Ben88}
{Bender}, R. 1988, \aap, 193, L7

\bibitem[{{Bender} {et~al.}(1992){Bender}, {Burstein}, \& {Faber}}]{BBF92}
{Bender}, R., {Burstein}, D., \& {Faber}, S.~M. 1992, \apj, 399, 462

\bibitem[{{Bender} {et~al.}(1988){Bender}, {Doebereiner}, \&
  {Moellenhoff}}]{BDM88}
{Bender}, R., {Doebereiner}, S., \& {Moellenhoff}, C. 1988, \aaps, 74, 385

\bibitem[{{Bender} \& {Nieto}(1990)}]{BN90}
{Bender}, R. \& {Nieto}, J.-L. 1990, \aap, 239, 97

\bibitem[{{Bender} {et~al.}(1989){Bender}, {Surma}, {Doebereiner},
  {Moellenhoff}, \& {Madejsky}}]{Ben89}
{Bender}, R., {Surma}, P., {Doebereiner}, S., {Moellenhoff}, C., \& {Madejsky},
  R. 1989, \aap, 217, 35

\bibitem[{{Bendo} \& {Barnes}(2000)}]{BB00}
{Bendo}, G.~J. \& {Barnes}, J.~E. 2000, \mnras, 316, 315

\bibitem[{{Binney}(1978)}]{Bin78}
{Binney}, J. 1978, \mnras, 183, 501

\bibitem[{{Binney}(1985)}]{Bin85}
---. 1985, \mnras, 212, 767

\bibitem[{{Binney} \& {Tremaine}(1987)}]{BT}
{Binney}, J. \& {Tremaine}, S. 1987, {Galactic dynamics} (Princeton, NJ,
  Princeton University Press, 1987, 747 p.)

\bibitem[{{Bournaud} {et~al.}(2004){Bournaud}, {Combes}, \& {Jog}}]{BCJ04}
{Bournaud}, F., {Combes}, F., \& {Jog}, C.~J. 2004, \aap, 418, L27

\bibitem[{{Bournaud} {et~al.}(2005){Bournaud}, {Jog}, \& {Combes}}]{BJC05}
{Bournaud}, F., {Jog}, C.~J., \& {Combes}, F. 2005, \aap, 437, 69

\bibitem[{{Boylan-Kolchin} {et~al.}(2005){Boylan-Kolchin}, {Ma}, \&
  {Quataert}}]{BK05}
{Boylan-Kolchin}, M., {Ma}, C.-P., \& {Quataert}, E. 2005, \mnras, 362, 184

\bibitem[{{Boylan-Kolchin} {et~al.}(2006){Boylan-Kolchin}, {Ma}, \&
  {Quataert}}]{BK06}
---. 2006, astro-ph/0601400

\bibitem[{{Burkert} \& {Naab}(2005)}]{BN05}
{Burkert}, A. \& {Naab}, T. 2005, \mnras, 363, 597

\bibitem[{{Capelato} {et~al.}(1995){Capelato}, {de Carvalho}, \&
  {Carlberg}}]{CdCC95}
{Capelato}, H.~V., {de Carvalho}, R.~R., \& {Carlberg}, R.~G. 1995, \apj, 451,
  525

\bibitem[{{Cappellari} {et~al.}(2006){Cappellari}, {Bacon}, {Bureau}, {Damen},
  {Davies}, {de Zeeuw}, {Emsellem}, {Falc{\'o}n-Barroso}, {Krajnovi{\'c}},
  {Kuntschner}, {McDermid}, {Peletier}, {Sarzi}, {van den Bosch}, \& {van de
  Ven}}]{Cap06}
{Cappellari}, M., {Bacon}, R., {Bureau}, M., {Damen}, M.~C., {Davies}, R.~L.,
  {de Zeeuw}, P.~T., {Emsellem}, E., {Falc{\'o}n-Barroso}, J., {Krajnovi{\'c}},
  D., {Kuntschner}, H., {McDermid}, R.~M., {Peletier}, R.~F., {Sarzi}, M., {van
  den Bosch}, R.~C.~E., \& {van de Ven}, G. 2006, \mnras, 366, 1126

\bibitem[{{Carlberg}(1986)}]{Car86}
{Carlberg}, R.~G. 1986, \apj, 310, 593

\bibitem[{{Cox} {et~al.}(2006){Cox}, {Di Matteo}, {Hernquist}, {Hopkins},
  {Robertson}, \& {Springel}}]{Cox06x}
{Cox}, T.~J., {Di Matteo}, T., {Hernquist}, L., {Hopkins}, P.~F., {Robertson},
  B., \& {Springel}, V. 2006, ApJ accepted (astro-ph/0504156)

\bibitem[{{Cox} {et~al.}(2005){Cox}, {Jonsson}, {Primack}, \&
  {Somerville}}]{Cox05}
{Cox}, T.~J., {Jonsson}, P., {Primack}, J., \& {Somerville}, R.~S. 2005, MNRAS,
  submitted (astro-ph/0503201)

\bibitem[{{Cox} {et~al.}(2004){Cox}, {Primack}, {Jonsson}, \&
  {Somerville}}]{Cox04}
{Cox}, T.~J., {Primack}, J., {Jonsson}, P., \& {Somerville}, R.~S. 2004, \apjl,
  607, L87

\bibitem[{{Cretton} {et~al.}(2001){Cretton}, {Naab}, {Rix}, \&
  {Burkert}}]{Cre01}
{Cretton}, N., {Naab}, T., {Rix}, H., \& {Burkert}, A. 2001, \apj, 554, 291

\bibitem[{{Davies} \& {Birkinshaw}(1988)}]{DB88}
{Davies}, R.~L. \& {Birkinshaw}, M. 1988, \apjs, 68, 409

\bibitem[{{Davies} {et~al.}(1983){Davies}, {Efstathiou}, {Fall}, {Illingworth},
  \& {Schechter}}]{Dav83}
{Davies}, R.~L., {Efstathiou}, G., {Fall}, S.~M., {Illingworth}, G., \&
  {Schechter}, P.~L. 1983, \apj, 266, 41

\bibitem[{{de Vaucouleurs}(1948)}]{deV48}
{de Vaucouleurs}, G. 1948, Annales d'Astrophysique, 11, 247

\bibitem[{{de Zeeuw} {et~al.}(2002){de Zeeuw}, {Bureau}, {Emsellem}, {Bacon},
  {Marcella Carollo}, {Copin}, {Davies}, {Kuntschner}, {Miller}, {Monnet},
  {Peletier}, \& {Verolme}}]{dZ02}
{de Zeeuw}, P.~T., {Bureau}, M., {Emsellem}, E., {Bacon}, R., {Marcella
  Carollo}, C., {Copin}, Y., {Davies}, R.~L., {Kuntschner}, H., {Miller},
  B.~W., {Monnet}, G., {Peletier}, R.~F., \& {Verolme}, E.~K. 2002, \mnras,
  329, 513

\bibitem[{{de Zeeuw} \& {Franx}(1991)}]{dZF91}
{de Zeeuw}, T. \& {Franx}, M. 1991, \araa, 29, 239

\bibitem[{{Di Matteo} {et~al.}(2005){Di Matteo}, {Springel}, \&
  {Hernquist}}]{dMSH05}
{Di Matteo}, T., {Springel}, V., \& {Hernquist}, L. 2005, \nat, 433, 604

\bibitem[{{Dubinski}(1998)}]{Dub98}
{Dubinski}, J. 1998, \apj, 502, 141

\bibitem[{{Erb} {et~al.}(2006){Erb}, {Steidel}, {Shapley}, {Pettini}, {Reddy},
  \& {Adelberger}}]{Erb06m}
{Erb}, D.~K., {Steidel}, C.~C., {Shapley}, A.~E., {Pettini}, M., {Reddy},
  N.~A., \& {Adelberger}, K.~L. 2006, ApJ accepted (astro-ph/0604041)

\bibitem[{{Faber} {et~al.}(1997){Faber}, {Tremaine}, {Ajhar}, {Byun},
  {Dressler}, {Gebhardt}, {Grillmair}, {Kormendy}, {Lauer}, \&
  {Richstone}}]{Fab97}
{Faber}, S.~M., {Tremaine}, S., {Ajhar}, E.~A., {Byun}, Y., {Dressler}, A.,
  {Gebhardt}, K., {Grillmair}, C., {Kormendy}, J., {Lauer}, T.~R., \&
  {Richstone}, D. 1997, \aj, 114, 1771

\bibitem[{{Faber} {et~al.}(1989){Faber}, {Wegner}, {Burstein}, {Davies},
  {Dressler}, {Lynden-Bell}, \& {Terlevich}}]{Fab89}
{Faber}, S.~M., {Wegner}, G., {Burstein}, D., {Davies}, R.~L., {Dressler}, A.,
  {Lynden-Bell}, D., \& {Terlevich}, R.~J. 1989, \apjs, 69, 763

\bibitem[{{Ferrarese} \& {Merritt}(2000)}]{FM00}
{Ferrarese}, L. \& {Merritt}, D. 2000, \apjl, 539, L9

\bibitem[{{Franx} {et~al.}(1991){Franx}, {Illingworth}, \& {de Zeeuw}}]{FIZ91}
{Franx}, M., {Illingworth}, G., \& {de Zeeuw}, T. 1991, \apj, 383, 112

\bibitem[{{Franx} {et~al.}(1989){Franx}, {Illingworth}, \& {Heckman}}]{FIH89}
{Franx}, M., {Illingworth}, G., \& {Heckman}, T. 1989, \apj, 344, 613

\bibitem[{{Gebhardt} {et~al.}(2000){Gebhardt}, {Bender}, {Bower}, {Dressler},
  {Faber}, {Filippenko}, {Green}, {Grillmair}, {Ho}, {Kormendy}, {Lauer},
  {Magorrian}, {Pinkney}, {Richstone}, \& {Tremaine}}]{Geb00}
{Gebhardt}, K., {Bender}, R., {Bower}, G., {Dressler}, A., {Faber}, S.~M.,
  {Filippenko}, A.~V., {Green}, R., {Grillmair}, C., {Ho}, L.~C., {Kormendy},
  J., {Lauer}, T.~R., {Magorrian}, J., {Pinkney}, J., {Richstone}, D., \&
  {Tremaine}, S. 2000, \apjl, 539, L13

\bibitem[{{Gerhard} \& {Binney}(1985)}]{GB85}
{Gerhard}, O.~E. \& {Binney}, J. 1985, \mnras, 216, 467

\bibitem[{{Gonz{\' a}lez-Garc{\'{\i}}a} \& {Balcells}(2005)}]{GG05a}
{Gonz{\' a}lez-Garc{\'{\i}}a}, A.~C. \& {Balcells}, M. 2005, \mnras, 357, 753

\bibitem[{{Gonz{\'a}lez-Garc{\'{\i}}a} \& {van Albada}(2003)}]{GG03}
{Gonz{\'a}lez-Garc{\'{\i}}a}, A.~C. \& {van Albada}, T.~S. 2003, \mnras, 342,
  L36

\bibitem[{{Gonz{\'a}lez-Garc{\'{\i}}a} \& {van Albada}(2005)}]{GG05c}
---. 2005, \mnras, 361, 1043

\bibitem[{{G{\'o}rski} {et~al.}(2005){G{\'o}rski}, {Hivon}, {Banday},
  {Wandelt}, {Hansen}, {Reinecke}, \& {Bartelmann}}]{Gor05}
{G{\'o}rski}, K.~M., {Hivon}, E., {Banday}, A.~J., {Wandelt}, B.~D., {Hansen},
  F.~K., {Reinecke}, M., \& {Bartelmann}, M. 2005, \apj, 622, 759

\bibitem[{{Hernquist}(1990)}]{H90}
{Hernquist}, L. 1990, \apj, 356, 359

\bibitem[{{Hernquist}(1992)}]{HRemI}
---. 1992, \apj, 400, 460

\bibitem[{{Hernquist}(1993)}]{HRemII}
---. 1993, \apj, 409, 548

\bibitem[{{Hernquist} \& {Barnes}(1991)}]{HB91}
{Hernquist}, L. \& {Barnes}, J.~E. 1991, \nat, 354, 210

\bibitem[{{Hernquist} \& {Spergel}(1992)}]{HS92}
{Hernquist}, L. \& {Spergel}, D.~N. 1992, \apjl, 399, L117

\bibitem[{{Hernquist} {et~al.}(1993){Hernquist}, {Spergel}, \&
  {Heyl}}]{HRemIII}
{Hernquist}, L., {Spergel}, D.~N., \& {Heyl}, J.~S. 1993, \apj, 416, 415

\bibitem[{{Heyl} {et~al.}(1994){Heyl}, {Hernquist}, \& {Spergel}}]{HRemIV}
{Heyl}, J.~S., {Hernquist}, L., \& {Spergel}, D.~N. 1994, \apj, 427, 165

\bibitem[{{Hopkins} {et~al.}(2005{\natexlab{a}}){Hopkins}, {Hernquist}, {Cox},
  {Di Matteo}, {Martini}, {Robertson}, \& {Springel}}]{Hop05b}
{Hopkins}, P.~F., {Hernquist}, L., {Cox}, T.~J., {Di Matteo}, T., {Martini},
  P., {Robertson}, B., \& {Springel}, V. 2005{\natexlab{a}}, \apj, 630, 705

\bibitem[{{Hopkins} {et~al.}(2005{\natexlab{b}}){Hopkins}, {Hernquist}, {Cox},
  {Di Matteo}, {Robertson}, \& {Springel}}]{Hop05c}
{Hopkins}, P.~F., {Hernquist}, L., {Cox}, T.~J., {Di Matteo}, T., {Robertson},
  B., \& {Springel}, V. 2005{\natexlab{b}}, \apj, 630, 716

\bibitem[{{Hopkins} {et~al.}(2005{\natexlab{c}}){Hopkins}, {Hernquist}, {Cox},
  {Di Matteo}, {Robertson}, \& {Springel}}]{Hop05d}
---. 2005{\natexlab{c}}, \apj, 632, 81

\bibitem[{{Hopkins} {et~al.}(2006{\natexlab{a}}){Hopkins}, {Hernquist}, {Cox},
  {Di Matteo}, {Robertson}, \& {Springel}}]{Hop06big}
---. 2006{\natexlab{a}}, \apjs, 163, 1

\bibitem[{{Hopkins} {et~al.}(2006{\natexlab{b}}){Hopkins}, {Hernquist}, {Cox},
  {Robertson}, {Di Matteo}, \& {Springel}}]{Hop06slope}
{Hopkins}, P.~F., {Hernquist}, L., {Cox}, T.~J., {Robertson}, B., {Di Matteo},
  T., \& {Springel}, V. 2006{\natexlab{b}}, \apj, 639, 700

\bibitem[{{Hopkins} {et~al.}(2006{\natexlab{c}}){Hopkins}, {Hernquist}, {Cox},
  {Robertson}, \& {Springel}}]{Hop06red}
{Hopkins}, P.~F., {Hernquist}, L., {Cox}, T.~J., {Robertson}, B., \&
  {Springel}, V. 2006{\natexlab{c}}, \apjs, 163, 50

\bibitem[{{Hopkins} {et~al.}(2005{\natexlab{d}}){Hopkins}, {Hernquist},
  {Martini}, {Cox}, {Robertson}, {Di Matteo}, \& {Springel}}]{Hop05a}
{Hopkins}, P.~F., {Hernquist}, L., {Martini}, P., {Cox}, T.~J., {Robertson},
  B., {Di Matteo}, T., \& {Springel}, V. 2005{\natexlab{d}}, \apjl, 625, L71

\bibitem[{{Kazantzidis} {et~al.}(2005){Kazantzidis}, {Mayer}, {Colpi}, {Madau},
  {Debattista}, {Wadsley}, {Stadel}, {Quinn}, \& {Moore}}]{Kaz05}
{Kazantzidis}, S., {Mayer}, L., {Colpi}, M., {Madau}, P., {Debattista}, V.~P.,
  {Wadsley}, J., {Stadel}, J., {Quinn}, T., \& {Moore}, B. 2005, \apjl, 623,
  L67

\bibitem[{{Kennicutt}(1998)}]{Kenn98}
{Kennicutt}, R.~C. 1998, \apj, 498, 541

\bibitem[{{Khochfar} \& {Burkert}(2001)}]{KB01}
{Khochfar}, S. \& {Burkert}, A. 2001, \apj, 561, 517

\bibitem[{{Kormendy} \& {Bender}(1996)}]{KB96}
{Kormendy}, J. \& {Bender}, R. 1996, \apjl, 464, L119+

\bibitem[{{Lake}(1989)}]{Lak89}
{Lake}, G. 1989, \aj, 97, 1312

\bibitem[{{Lambas} {et~al.}(1992){Lambas}, {Maddox}, \& {Loveday}}]{LML92}
{Lambas}, D.~G., {Maddox}, S.~J., \& {Loveday}, J. 1992, \mnras, 258, 404

\bibitem[{{Lauer} {et~al.}(2005){Lauer}, {Faber}, {Gebhardt}, {Richstone},
  {Tremaine}, {Ajhar}, {Aller}, {Bender}, {Dressler}, {Filippenko}, {Green},
  {Grillmair}, {Ho}, {Kormendy}, {Magorrian}, {Pinkney}, \& {Siopis}}]{Lau05}
{Lauer}, T.~R., {Faber}, S.~M., {Gebhardt}, K., {Richstone}, D., {Tremaine},
  S., {Ajhar}, E.~A., {Aller}, M.~C., {Bender}, R., {Dressler}, A.,
  {Filippenko}, A.~V., {Green}, R., {Grillmair}, C.~J., {Ho}, L.~C.,
  {Kormendy}, J., {Magorrian}, J., {Pinkney}, J., \& {Siopis}, C. 2005, \aj,
  129, 2138

\bibitem[{{Lidz} {et~al.}(2006){Lidz}, {Hopkins}, {Cox}, {Hernquist}, \&
  {Robertson}}]{Lidz06}
{Lidz}, A., {Hopkins}, P.~F., {Cox}, T.~J., {Hernquist}, L., \& {Robertson}, B.
  2006, \apj, 641, 41

\bibitem[{{Merritt}(1985)}]{Mer85}
{Merritt}, D. 1985, \apj, 289, 18

\bibitem[{{Merritt} \& {Fridman}(1996)}]{MF96}
{Merritt}, D. \& {Fridman}, T. 1996, \apj, 460, 136

\bibitem[{{Merritt} \& {Tremblay}(1996)}]{MT96}
{Merritt}, D. \& {Tremblay}, B. 1996, \aj, 111, 2243

\bibitem[{{Mihos} \& {Hernquist}(1994{\natexlab{a}})}]{MH94dsc}
{Mihos}, J.~C. \& {Hernquist}, L. 1994{\natexlab{a}}, ApJL, 437, L47

\bibitem[{{Mihos} \& {Hernquist}(1994{\natexlab{b}})}]{MH94popg}
---. 1994{\natexlab{b}}, \apj, 427, 112

\bibitem[{{Mihos} \& {Hernquist}(1994{\natexlab{c}})}]{MH94majm}
---. 1994{\natexlab{c}}, \apjl, 431, L9

\bibitem[{{Mihos} \& {Hernquist}(1996)}]{MH96}
---. 1996, \apj, 464, 641

\bibitem[{{Milosavljevi{\' c}} \& {Merritt}(2001)}]{Mil01}
{Milosavljevi{\' c}}, M. \& {Merritt}, D. 2001, \apj, 563, 34

\bibitem[{{Naab} \& {Burkert}(2003)}]{NB03}
{Naab}, T. \& {Burkert}, A. 2003, \apj, 597, 893

\bibitem[{{Naab} {et~al.}(2006){Naab}, {Khochfar}, \& {Burkert}}]{NKB06}
{Naab}, T., {Khochfar}, S., \& {Burkert}, A. 2006, \apjl, 636, L81

\bibitem[{{Negroponte} \& {White}(1983)}]{NW83}
{Negroponte}, J. \& {White}, S.~D.~M. 1983, \mnras, 205, 1009

\bibitem[{{Nipoti} {et~al.}(2003){Nipoti}, {Londrillo}, \& {Ciotti}}]{NLC03}
{Nipoti}, C., {Londrillo}, P., \& {Ciotti}, L. 2003, \mnras, 342, 501

\bibitem[{{Ostriker}(1980)}]{Ost80}
{Ostriker}, J.~P. 1980, Comments on Astrophysics, 8, 177

\bibitem[{{Rix} {et~al.}(1999){Rix}, {Carollo}, \& {Freeman}}]{RCF99}
{Rix}, H., {Carollo}, C.~M., \& {Freeman}, K. 1999, \apjl, 513, L25

\bibitem[{{Robertson} {et~al.}(2006{\natexlab{a}}){Robertson}, {Cox},
  {Hernquist}, {Franx}, {Hopkins}, {Martini}, \& {Springel}}]{Rob06fp}
{Robertson}, B., {Cox}, T.~J., {Hernquist}, L., {Franx}, M., {Hopkins}, P.~F.,
  {Martini}, P., \& {Springel}, V. 2006{\natexlab{a}}, \apj, 641, 21

\bibitem[{{Robertson} {et~al.}(2005){Robertson}, {Hernquist}, {Bullock}, {Cox},
  {Di Matteo}, {Springel}, \& {Yoshida}}]{Rob05a}
{Robertson}, B., {Hernquist}, L., {Bullock}, J.~S., {Cox}, T.~J., {Di Matteo},
  T., {Springel}, V., \& {Yoshida}, N. 2005, arXiv:astro-ph/0503369

\bibitem[{{Robertson} {et~al.}(2006{\natexlab{b}}){Robertson}, {Hernquist},
  {Cox}, {Di Matteo}, {Hopkins}, {Martini}, \& {Springel}}]{Rob06b}
{Robertson}, B., {Hernquist}, L., {Cox}, T.~J., {Di Matteo}, T., {Hopkins},
  P.~F., {Martini}, P., \& {Springel}, V. 2006{\natexlab{b}}, \apj, 641, 90

\bibitem[{{Rothberg} \& {Joseph}(2004)}]{RJ04}
{Rothberg}, B. \& {Joseph}, R.~D. 2004, \aj, 128, 2098

\bibitem[{{Rothberg} \& {Joseph}(2006)}]{RJ06}
---. 2006, \aj, 131, 185

\bibitem[{{Ryden}(1992)}]{Ryd92}
{Ryden}, B. 1992, \apj, 396, 445

\bibitem[{{Schmidt}(1959)}]{Sch59}
{Schmidt}, M. 1959, \apj, 129, 243

\bibitem[{{Schweizer}(1998)}]{Schweizer98}
{Schweizer}, F. 1998, in Saas-Fee Advanced Course 26: Galaxies: Interactions
  and Induced Star Formation, ed. D.~{Friedli}, L.~{Martinet}, \&
  D.~{Pfenniger}

\bibitem[{{Sersic}(1968)}]{Ser68}
{Sersic}, J.~L. 1968, {Atlas de galaxias australes} (Cordoba, Argentina:
  Observatorio Astronomico, 1968)

\bibitem[{{Shen} {et~al.}(2003){Shen}, {Mo}, {White}, {Blanton}, {Kauffmann},
  {Voges}, {Brinkmann}, \& {Csabai}}]{Shen03}
{Shen}, S., {Mo}, H.~J., {White}, S.~D.~M., {Blanton}, M.~R., {Kauffmann}, G.,
  {Voges}, W., {Brinkmann}, J., \& {Csabai}, I. 2003, \mnras, 343, 978

\bibitem[{{Springel}(2000)}]{Sp00}
{Springel}, V. 2000, \mnras, 312, 859

\bibitem[{{Springel}(2005)}]{SpGad2}
---. 2005, \mnras, 364, 1105

\bibitem[{{Springel} {et~al.}(2005{\natexlab{a}}){Springel}, {Di Matteo}, \&
  {Hernquist}}]{SdMH05red}
{Springel}, V., {Di Matteo}, T., \& {Hernquist}, L. 2005{\natexlab{a}}, \apjl,
  620, L79

\bibitem[{{Springel} {et~al.}(2005{\natexlab{b}}){Springel}, {Di Matteo}, \&
  {Hernquist}}]{SdMH05}
---. 2005{\natexlab{b}}, \mnras, 361, 776

\bibitem[{{Springel} \& {Hernquist}(2002)}]{SHEnt}
{Springel}, V. \& {Hernquist}, L. 2002, \mnras, 333, 649

\bibitem[{{Springel} \& {Hernquist}(2003)}]{SH03}
---. 2003, \mnras, 339, 289

\bibitem[{{Toomre}(1977)}]{T77}
{Toomre}, A. 1977, in Evolution of Galaxies and Stellar Populations, p.401

\bibitem[{{Toomre} \& {Toomre}(1972)}]{TT72}
{Toomre}, A. \& {Toomre}, J. 1972, \apj, 178, 623

\bibitem[{{Tremblay} \& {Merritt}(1995)}]{TM95}
{Tremblay}, B. \& {Merritt}, D. 1995, \aj, 110, 1039

\bibitem[{{Trujillo} {et~al.}(2001){Trujillo}, {Aguerri}, {Cepa}, \&
  {Guti{\'e}rrez}}]{Tru01}
{Trujillo}, I., {Aguerri}, J.~A.~L., {Cepa}, J., \& {Guti{\'e}rrez}, C.~M.
  2001, \mnras, 321, 269

\bibitem[{{van Dokkum}(2005)}]{vD05}
{van Dokkum}, P.~G. 2005, \aj, 130, 2647

\end{thebibliography}


\end{document}